%
\magnification=\magstep1
\def\oneandahalfspace{\baselineskip=\normalbaselineskip
  \multiply\baselineskip by 3 \divide\baselineskip by 2}

\catcode`\@=11
\def\vereq#1#2{\lower3pt\vbox{\baselineskip1.5pt \lineskip1.5pt
\ialign{$\m@th#1\hfill##\hfil$\crcr#2\crcr\sim\crcr}}}
\catcode`\@=12
\overfullrule=0pt
 
\def\ref{\hangindent=50pt \hangafter=1 \noindent}
\def\etal {{\it {et al}\/}\ }

\def\deg{$^\circ$}

\nopagenumbers
\null
\bigskip
\centerline{\bf THE SLOAN DIGITAL SKY SURVEY PHOTOMETRIC CAMERA}

%
%
\bigskip
\centerline{J. E. Gunn$^{1\dagger}$, M. Carr$^1$, C. Rockosi$^3$, M. Sekiguchi$^2$}
\bigskip
\centerline{K. Berry, B. Elms, E. de Haas, Z. Ivezic, G. Knapp, R. Lupton, G. Pauls, R. Simcoe}
\centerline{Princeton University Observatory}
\medskip
\centerline{R. Hirsch, D. Sanford, S. Wang, D. York}
\centerline{Department of Astronomy and Astrophysics, University of Chicago}
\medskip
\centerline{F. Harris}
\centerline{U. S. Naval Observatory, Flagstaff}
\medskip
\centerline{J. Annis, L. Bartozek, W. Boroski, J. Bakken, M. Haldeman, S. Kent, }
\centerline{S. Holm, D. Holmgren, D. Petravick, A. Prosapio, R. Rechenmacher}
\centerline{Fermi National Accelerator Laboratory}
\medskip
\centerline{M. Doi$^4$, M. Fukugita$^{2,5}$, K. Shimasaku$^4$}
\centerline{University of Tokyo}
\medskip
\centerline{N. Okada}
\centerline{National Astronomical Observatory of Japan}
\medskip
\centerline{C. Hull, W. Siegmund, E. Mannery}
\centerline{Department of Astronomy, University of Washington}
\medskip
\centerline{M. Blouke, D. Heidtman}
\centerline{Scientific Imaging Technologies}
\medskip
\centerline{D. Schneider}
\centerline{Department of Astronomy, Pennsylvania State University}
\medskip
\centerline{R. Lucinio, J. Brinkman}
\centerline{Apache Point Observatory}
\medskip
\vfil
\centerline{$^1$ Princeton University Observatory, Princeton, NJ 08544}
\centerline{$^2$ Institute for Cosmic Ray Research, University of Tokyo}
\vskip -0.5mm
\centerline{Tanashi, Tokyo 188, Japan}
\centerline{$^3$ Department of Astronomy and Astrophysics, University of 
 Chicago, Chicago, IL60637}
\centerline{$^4$ Department of Astronomy, University of Tokyo}
\vskip -0.5mm
\centerline{Hongo, Tokyo 113, Japan}
\centerline{$^5$ Institute for Advanced Study, Princeton, NJ 08540}
\medskip
\centerline{$^\dagger$Electronic mail:~ jeg@astro.princeton.edu}
\eject 
\footline={\hss\tenrm\folio\hss}
\noindent
\oneandahalfspace
{\bf Abstract}
\medskip
We have constructed a large format mosaic CCD camera for the
Sloan Digital Sky Survey. The camera consists of two arrays, a photometric
array which uses 30 2048$\times$2048 SITe/Tektronix CCDs 
(24 micron pixels) with
an effective imaging area of 720 cm$^2$,
and an astrometric array which uses 24 400$\times$2048 CCDs with the same
pixel size which will allow us to
tie bright astrometric standard stars to the objects imaged in the
photometric camera. The instrument will be used to carry out photometry
essentially simultaneously in five color bands spanning the range
accessible to silicon detectors on the ground in the
time--delay--and--integrate (TDI) scanning mode. The photometric 
detectors are arrayed in the focal plane in six columns of five chips 
each such that two scans cover a filled stripe 2.5 degrees wide.
This paper presents engineering and technical details of the camera.
\vfil\eject

\noindent
{\bf 1. Introduction}\hfil

The Sloan Digital Sky Survey (SDSS) is a project undertaking a digitized 
photometric survey of half the northern sky to about 23rd mag and a follow--up
spectroscopic survey of one million galaxies and 100,000 quasars
within precisely defined selection criteria. This project aims at
producing a large homogeneous data sample of the northern sky with
high accuracy multi--color photometry and accurate astrometry.
A wide--field telescope with accurate and well--understood astrometric
properties and very precise drives, a large format imaging camera, 
a high multiplex gain multiobject
spectrograph, and a data and software system capable of dealing with the
high data rates and voluminous data
are the indispensable elements for this project.

For the camera, we have constructed a large array of large format
CCD detectors mosaicked on the focal plane of a dedicated 2.5 meter
telescope with a new modified distortion--free Ritchey--Chr\'etien optical design.
Our requirements for optical performance are quite demanding:
for a large--area survey, it is crucial to realize a wide field of
view for efficient photometric imaging, and fast final focal ratios
for reasonable telescope apertures in order to match the pixel
size for available CCD detectors . It is known that a wide, flat 
field of view can be achieved with
a  Ritchey--Chr\'etien design with two hyperboloids with the same curvature
(Bowen \& Vaughan 1973). The most successful telescope of this type
is the Swope 1 m at Las Campanas, which has a 2.9\deg~ field of view
(FoV) with a final focal ratio of f/7.
The SDSS telescope has a diameter of 2.5m with
a faster final focal ratio f/5 and a slightly larger field (3\deg). 

Our demands for optical performance are substantially
more difficult to realize than are typically required in astronomical systems.
We plan to carry out TDI
drift scans in the imaging survey in order to increase the observing
efficiency. TDI saves both the read--out time
of the CCD, which would be comparable to the exposure time in ordinary
``snapshot'' mode, and repointing and settling time. We will also thereby 
achieve very good flat--fielding, because the flat field becomes a 
one--dimensional vector; defects along a column are averaged over. 
This imaging technique requires
carefully controlled distortion, since either
a change of scale or a differential deviation from conformality
of the mapping of the sky onto the focal plane across a chip
immediately causes image degradation. Distortion is not normally an
aberration of much concern to astronomers, and the classical optical
systems used in astronomy are generally not very well corrected for it;
in particular, the Gascoigne corrector used
to correct astigmatism in the usual Ritchey arrangement introduces
distortion which is more than an order of magnitude larger than would
be acceptable for our configuration. 

We therefore developed a new Ritchey--like design using {\it two} aspheric
corrector lenses, one of the classical Gascoigne Schmidt--plate shape
about 75 cm in front of the focal plane and the other, quite thick, with
a negative Schmidt--plate form and much more strongly figured, just in
front of the focal plane.  This combination works by virtue of the fact
that the astigmatism introduced by a Gascoigne plate is proportional to
its aspheric amplitude and the square of the distance from the focal
plane, whereas the distortion is proportional to the amplitude and the
first power of the distance from the focal plane.  Thus two correctors
of opposite sign and different distances from the focus can both
accurately remove astigmatism and distortion.  It is fortuitous that the
same combination also removes residual lateral color, which has the same
dependence on the parameters as distortion.  The focal surface is almost
flat, deviating from a plane by less than 1.3 mm over the full 3 degree
(650 mm) field.  The optical design, its optimization for this
configuration, and the desiderata which led to it will be described more
fully in another publication. 

The efficiency of a large--area survey is characterized by the quantity
$$\epsilon=\Omega D^2 q,\eqno{(1)}$$
where $\Omega$ is the solid angle of the FoV, $D$ the diameter of the
telescope and $q$ is the detective quantum efficiency on the sky, 
assuming that the seeing disk is resolved; all else being equal, the
time to completion to a given depth depends inversely on $\epsilon$.
The SDSS attains $\epsilon=4.8$ m$^2$deg$^2$, which is compared to
$\approx 0.3$ m$^2$deg$^2$ for 48$''$ Schmidt photography.
This gain enables us both to go significantly deeper than existing
photographic surveys and to carry out imaging simultaneously
in five color bands which cover the wavelength range from the atmospheric
cutoff in the ultraviolet to the silicon cutoff in the infrared. The
bands we have chosen are
\line{\hfil}
\line{$u'$ ($\lambda_{\rm eff} = 3550$\AA{}), \hfil}
\line{$g'$ ($\lambda_{\rm eff} = 4770$\AA{}). \hfil}
\line{$r'$ ($\lambda_{\rm eff} = 6230$\AA{}), \hfil}
\line{$i'$ ($\lambda_{\rm eff} = 7620$\AA{}), \hfil}
\line{$z'$ ($\lambda_{\rm eff} = 9130$\AA{}). \hfil}
\line{\hfil}
This is a new photometric system to optimize the
physical advantages of each color band appropriate for the
galaxy survey. Further details of the photometric system are
available in Fukugita \etal (1996).

The system design was based on the then--largest--available CCD detectors,
the Tektronix (later SITe) Tk2048E, which have a 2048$\times$2048 array
of 24 $\mu$m pixels.  The focal length of the telescope is designed to
match this device; the image scale 16.$''$5 mm$^{-1}$ corresponds to
0.4$''$ per 24 micron pixel.  With the 0.8$''$ median free--air seeing 
measured at
the Apache Point Observatory site, one expects about 1.0$''$ FWHM images
taking into account other items in the error budget.  This represents
quite excellent sampling, with less than 1 per cent of the power in a
star beyond the Nyquist frequency. 

We can accommodate in our 3 degree field a 5(colors)$\times$6(columns)
array of 2048$\times$2048 chips with an appropriate spacing among the
chips; the active area of the devices is 49.15 mm (13.52$'$), the CCD
packages are 63.5 mm square, and the center--to--center spacing along a
column is 65mm, about 18.0$'$.  The center--to--center spacing between
columns is 91 mm, about 25.2$'$, slightly less than twice the active
width of the chips, so that two scans cover a filled stripe 2.54\deg
wide, with an 8\% (about 1$'$) overlap between the two scans along each
edge.  We will scan at sidereal rate, so the effective exposure time is
54.1 seconds for each of the 5 colors along a column; a star remains on
the photometric array for 342 seconds, and the spacing in time from one
color to the next is 72 seconds.  The limiting magnitudes (defined to be
at S/N=5) in an AB system for the five colors are expected to be about
22.1 for $u'$, 23.2 for $g'$, 23.1 for $r'$, 22.5 for $i'$, and 20.8 for
$z'$ for stellar images at an airmass of 1.4, near the planned median for the
survey.

We use the extra space in the focal plane above and below the
photometric array to arrange 22 smaller CCD chips (2048$\times$400 with
24 $\mu$m pixels) for astrometry and two more as monitors for automated
focusing.  These are arrayed (see Fig.  1) in a dewar with twelve
devices above (the {\it leading} astrometric array) the photometric
array and another (the {\it trailing} astrometric array) identical array
of 12 below the main array.  The eleven astrometric devices in one dewar
are arranged
in two rows, one of six aligned with the photometric columns (the
{\it column chips}) and a second of five (the {\it bridge chips})
straddling the columns.  The centers of the first row and second
row of astrometric devices are 204.5 and 220 mm, respectively, above
and below the center of the field.
The astrometric CCDs have passband filters
nominally identical to the $r'$ ones and 3.0 magnitudes of neutral
density filters as well.  Data from the astrometric array enables us to
tie the coordinates of the objects observed in the photometric camera to
the reference astrometric system which is based on bright stars that
would saturate the photometric camera in the survey mode.  The choice of
filters for the astrometric devices is such that there is a roughly
three--magnitude range of useful overlap between them and the photometric
array.  The two focus devices, whose centers are 235.5 mm above and below
the center, have $r'$ filters as well, though no
neutral filters, and have a three--piece stepped thickness plate
incorporated into the filter.  Though they are pretty far out in the
field, the optical design (largely fortuitously) delivers excellent
images there at the relevant wavelength, and one can do very critical
focusing. The trailing astrometric array not only increases the
astrometric precision by a factor of $\sqrt 2$ but allows accurate
monitoring of the tracking of the telescope.  

Though we believe that this camera is the largest and most complex of
its kind now extant and operating, there are many other cameras
operating, near operation, or under construction which are at similar
levels of complexity.  The NAO/UTokyo mosaic, (Sekiguchi \etal 1992,
Kashikawa \etal 1995) based on 1K $\times$ 1K TIJ virtual-phase devices,
has been in operation in various guises for several years and now is
using 40 CCDs, though they cannot be read simultaneously.  The MOA
mosaic (Abe \etal 1997) is a smaller array using these chips.  Another
large array using relatively small devices is the ESO one used for
microlensing searches (Arnaud \etal 1994).  Tyson's group has built and has
been using for some years a camera with 4 of the 2K $\times$ 2K devices
like ours (Wittman \etal 1998).
Most of the attention now is
on mosaics of 3-side buttable 2K $\times$ 4K devices with 15 micron
pixels being produced or promised by several manufacturers, including
SITe, Lincoln Laboratories, Orbit, EEV, Hammamatsu, and Loral.  Cameras
using these devices either in operation or in advanced stages of testing
have been described by Abe \etal (in preparation, 3 $\times$ 1), Ives
\etal (1996, 4 $\times$ 1), Luppino \etal (1996, 4 $\times$ 2), and Boroson
\etal (1994, 4 $\times$ 2).  Another relatively large pair of arrays,
the EROS cameras,
each using eight 2K $\times$ 2K Loral devices are described in Bauer and
DeKat (1998).  Even larger cameras are planned using the 2K $\times$ 4K
devices, including a 2 $\times$ 5 array for Subaru (Miyazaki \etal 1998)
and the CFHT ``MEGACAM'' project (Boulade 1998), using an 8 $\times$ 4
array, which will be larger in terms of total pixels than the SDSS
camera. Two large mosaics of these chips will be used in the DEIMOS
spectrograph for the Keck telescope (James \etal 1998).

The field of our system is so large that one must perform the 
TDI scans along great
circles in order to obtain satisfactory image quality, but the ability
to park the telescope on the equator and take data is perceived by all
to be a great advantage both for testing and for obtaining the highest
astrometric accuracy, and was a significant driver in the choice of the
scanning rate.  The SDSS will make use of this in our Southern Survey,
in which a stripe one full camera width (2.5 degrees) wide and about 90
degrees long on the equator will be imaged in this fashion
repeatedly when the northern Galactic cap is inaccessible. 

We are aiming for high astrometric accuracy: Kolmogorov seeing
theory with parameters relevant to our site suggests that we should be
limited by seeing at about the 30-40 milliarcsec (mas) level, and we have
striven to be in a position so that seeing will {\it be} the limiting
factor in astrometric performance, both in the design and construction
of the camera and the telescope and in the design, specification, and
construction of the telescope drives and mirror supports and controls. 
Thirty mas corresponds to 2 $\mu$m
in the focal plane, and stability at this level is not trivial to achieve
over such a large focal plane.

Thus the problem of the design of the camera comes down to housing the
54 detectors in a way which is geometrically stable at the few micron
level, adjustable to conform to the focal surface, allows them to work
cooled to CCD operating temperature (about -80\deg C) in a good vacuum,
and attend to their complex electronic needs with sufficiently
modular and compact circuitry that assembly and maintenance are not an
impossible nightmare.  This has been altogether a rather challenging
set of problems, and in this paper we discuss how they have been solved. 

The plan of this paper is as follows. 
In section 2 we briefly review the telescope optics insofar as they
are relevant to the camera. In section
3 we discuss the photometric system and the characteristics of
the CCD chips. Section 4 describes the mechanical design of the 
photometric array, and section 5 the astrometric array and focus monitor.
There is a discussion of the CCD cooling system in section 6, and the
electronics is detailed in section 7.
The overall mechanical structure of the camera, its ``life--support'' system,
and its mounting to the telescope is discussed in section 8.

\medskip
\noindent
{\bf 2. Telescope Optics: Design and Performance}\hfil

We have reviewed the principles of the optical design in the introduction;
here we concentrate only on those details which are relevant to the design
of the camera, namely the final distortion corrector, the form of the focal
surface, and the imaging performance.

The focal surface sagitta $s$(mm) is given adequately by 
$$ 
s = -0.276 + 2.754\times 10^{-5} r^2 - 4.724\times 10^{-10} r^4 + 
        2.870\times 10^{-16} r^6
$$
where $r$ is the field radius in mm.  The design is almost distortion--free
in the sense that the radius in the focal plane is proportional, to high
accuracy, to the field angle (not its sine or tangent); zero distortion
for most wide--field imaging is defined for the condition that the radius
in the focal plane is proportional to the {\it tangent} of that angle,
which results in faithful representations of figures on {\it planes},
but we wish as faithfully as possible to image figures on a sphere onto
a surface which is almost planar.  For this case a compromise is
necessary between the wishes for constant scale in the sense that
meridians have constant linear separation in the focal plane, and the
desire that parallels of latitude do likewise.  The optimum case depends
somewhat on the aspect ratio of the field and is somewhere between the
radius in the focal plane going like the sine of the input angle and its
tangent.  For a square focal plane, which is close to the situation at
hand, the radius approximately proportional to the angle itself is the
best, and we have made this choice.  The errors can be minimized by
clocking different chips at different rates to correspond to the local
scale along the columns, but we have chosen not to do so for reasons of
noise reduction and simplicity in the data system.  Our design results
for the best compromise tracking rate in worst--case image smearing along
the columns of 0.06 arcseconds, 3 microns, or 0.14 pixels over the
imaging array.  Stars do not quite follow straight trajectories in the
focal plane, but this can be compensated for by a slight rotation of the
outer chips, amounting to about 0.006 degrees at the corners.  If it is
ignored, the resulting error is about 0.24 pixel; both of these are
entirely negligible (and are, in fact, of the same order as residual
distortions arising from the deviations of the focal height from the
desired strictly linear relationship with angle), but the problem quickly
becomes severe for bigger fields and can only be properly addressed
with anamorphic optics. 

The second corrector is very close to the focal plane, and we made use
of that in a very fundamental way in the design of the camera: we use
this element, which is made of fused silica, as the substrate
upon which the camera is built, and rely upon it to maintain the exacting
mechanical tolerances required for image quality and astrometry. It has a
central thickness of 45mm and the front face is very strongly aspheric, having 
an aspheric sagitta of more than 8 mm. The required global accuracy, however,
is not terribly stringent since it is placed close to the focal plane.
It was figured using mechanical metrology by Loomis Custom Optics to
a set of specifications which basically placed limits on the structure
function of the slope of the element to assure negligible image degradation
and astrometric error. The rear (plane) face of the corrector is 13mm
above the focal surface at the center, about 10mm at the extreme edge. The
thickness was chosen primarily for mechanical strength and stiffness
in view of the mechanical role it plays, with some small detriment to the 
image quality owing to the longitudinal color such a thick element introduces.
This plane face is the surface to which the dewars which house the CCDs and
the kinematic mounts for the optical benches upon which the CCDs are mounted
are attached and registered.

The CCDs are mounted in such a way that they can be adjusted to conform
to the focal surface.  This requires a tilt slightly smaller than a
degree at the edge of the field.  There is one further complication
brought about by the fact that the CCDs as produced are slightly convex,
with a reasonably well controlled radius of about 2.2 meters.  Thus the
best fit plane results in focus errors of about 100$\mu$m rms, which at
f/5 corresponds to an image degradation of about 20$\mu$m.  We correct
this curvature (to the mean chip radius---corrections for each chip
individually results in unacceptable scale variations from device to
device) for each chip with weak field
deflatteners cemented to the rear face of the filter, which in turn is
cemented to the backside of the second corrector surface.  The central
thickness of the filter/deflattener element is 5mm for the photometric
filters and 6mm for the astrometric ones, so the vertices of the CCDs
are nominally about 8 and 7 mm behind the filters. 

The front side of the second corrector is antireflection coated in four
strips that match each color band (the same coating was used for $i'$ and $z'$)
using appropriate masks in the coating process.  The coating was
done by QSP Optical Technologies and results in reflectivities below
0.2\% in each band. Thus there are only two surfaces near the detectors,
and, with the excellent antireflection coatings on the corrector, 
the primary source of ghosts in this system 
is reflection from the CCD surface to the back surface of the filters only about
7 mm away and
back. Though the interference coatings of the filters (which are on this
back surface) are quite good antireflection coatings in band, there are
inevitable very high reflectivities (accompanied by very low transmission)
in the short--pass cutoff region, and
at the cutoff wavelength the transmission {\it and} reflectance are necessarily
of order 50\%. It is in these narrow transition spectral regions that most
of the ghost flux originates; the $u'$ and $z'$ do not suffer from this
phenomenon because they do not use interference short--pass filters,
but the others do. (The filters and their makeup
are discussed in more detail in Fukugita \etal 1996.)

The discussion of the optical performance of the camera configuration is
a bit complicated because of the complexity of the focal plane, with
different filters in different locations and the
effect of residual distortion on the final TDI image quality. 

To facilitate more detailed discussion of image quality we show the
optical layout of the camera focal plane in Figure 1, which shows the 
locations of the
30 2048$\times$2048 photometric CCDs, the 22 2048$\times$400 astrometric
chips, and the two 2048$\times$400 focus--monitoring sensors.  The five filters
are arranged in (temporal, leading to trailing) order along the columns:
$r'$, 
$i'$,
$u'$, 
$z'$, and 
$g'$.

The camera is right--left reflection symmetric and the lower 
astrometric/focus array
is the mirror image of the upper array,  
so only 22 chips are given identifying field numbers.
The direction of the TDI scan
is upwards in this diagram, so a given star first encounters an
upper (leading) astrometric device, then an $r'$ chip, then a 
$i'$ chip, and so on until, 485
seconds later, it encounters the trailing astrometric chip located at the
bottom in the figure.  The arrow points to the extreme field radius 
used by the camera. This is at the corner of field 18, 327.6mm or 
about 90.6 arcminutes from the center.

To evaluate the image quality, we have performed a polychromatic
raytrace and composited several images along a CCD column to simulate
the effects of the TDI scan.  For each CCD, at each point of the
$5 \times 5$ array on the device 
the system has been traced with five
wavelengths chosen such that each is the mean wavelength of the
corresponding quintile of the filter/system photon response; thus each
has equal weight in the final image. 

The final images (five per CCD) are composed of the five individual
monochromatic images and, because TDI integrates along a column, of a 
composite of the
five images along a CCD column, taking account in the first instance of
any lateral color shifts and in the second of any residual distortion
perpendicular to the column and residual distortion and scale error
along the column.  The images are defocused to lie in the best--fitting
focal surface with the mean curvature of the CCDs for each subfield
(tilt and piston are fitted).  The input angles along the column
accurately represent images at successive equally spaced time intervals
in TDI mode, and the geometry on the sky for TDI is accurately modeled. 
The final images are shown in Figure 2. There are two panels: 
the bottom panel
shows the images as delivered by the design optical system, and  the top as
convolved with 0.8 arcsecond (FWHM) Gaussian seeing.
The PSFs were generated
by fitting discrete Zernike polynomials to the slope errors in the
system and using those fits to generate intercepts in the desired focal
plane for 1200 rays for each of the 25 images which go into the
polychromatic TDI composite.  Those rays were binned in 0.05
arcsecond pixels to generate the intensities for the grey scale images. 
In each panel, each row of
images is the model of TDI output for the array as labeled in
Figure 1; thus the bottom row consists of five images each across 
fields 13, 14, and 15,
the next 7, 8, and 9, and so forth.  The top two rows are the
astrometric fields 16, 17, 18, and 19, 20, 21.  The images for the focus
array are discussed and shown later in section 5.  
The spacing between successive
closely spaced images in the mosaic is 3 arcseconds.  

The situation is summarized quantitatively in Table 1,
where each row
lists the properties of one detector field. 
The table lists the
field center
in millimeters measured from the optical axis (--y is the TDI scan
direction), the size of the CCD for that field, the filter, the field
flattener curvature in units of $10^{-3}{\rm mm}^{-1}$ (ffc3), the CCD
curvature in units of $10^{-4}{\rm mm}^{-1}$ (ccd4), the vertical (along the
CCD columns, the scanning direction) scale
in that field (vscl in mm/arcmin), the rms focus error in microns over the CCD
caused by mismatch between the final best focal surface and the curved
CCD surface, the residual field curvature in units of $10^{-4} {\rm
mm}^{-1}$ (dc4), and the minimum (em) and maximum (eM) rms image diameters over
the field in microns.  We should perhaps comment on the residual
curvature; the overall scale in the focal surface is 3.615 mm (arcmin)$^{-1}$, 
for the optical design, 
but the field deflatteners change the scale locally for each chip to a
number close to 3.643, which is the ``scan scale'', i.e., the assumed
tracking rate.  Changes in this scale from chip to chip, and color
to color, represent errors in the TDI images, and the field flattener
curvatures are chosen for the best compromise between keeping the scale
constant and matching the focal curvature. The as-manufactured optics give
a scale very nearly the design value, but uncertain by about a unit in the
third place; in practice, the scale will be determined by tests on the sky.
Scale errors are in general
much more serious for image quality than focal errors, so there is
usually some residual curvature. 

The results indicate that for the photometric array, the maximum
rms image diameters are for the ultraviolet fields, reaching 0.63
arcseconds for the outermost one. The difficulty of achieving good
images in the $u'$ band (owing to the loss of proper aberration 
correction by the correctors with the large index shift at $u'$)
is the reason we have placed the $u'$ row in the center. 
The increase from 27 microns, the average diameter of monochromatic
non--TDI images in the $u'$ field 3, to 38 microns 
in the full TDI polychromatic treatment is mostly due to longitudinal
color, with tiny contributions from defocus, lateral color, and substantial
ones from
TDI effects.  Images as large are seen at the field extremes 
at the other end of the spectrum in $z'$, where they reach 39 microns,
0.65 arcsec rms. (Recall that for gaussian images, the FWHM is 0.83 times
the rms diameter; these aberrated images are not by any means gaussian,
but if the seeing dominates the image size, the resulting convolution
does not change the form of the PSF very much, and for image degradation
considerations these worst images can be considered equivalent to gaussians
with FWHMs of 0.53$''$  for the $u'$ and 0.54$''$ for the $z'$ one.
The other images are of order 0.5 arcsec or better (rms) over the whole field.
The problems in $z'$ are also due to the extreme
wavelength; the optimization of the system involves balancing the color
effects at the wavelength extremes, and because the polychromatic effects
at $u'$ are so large the {\it monochromatic} optimization is biased 
toward the ultraviolet. The optimization of the system was done in a
very detailed way, taking account of the filter/detector combination at
a given location in the focal plane. There is, of course, further
image degradation caused by differential atmospheric refraction, which
is fairly serious in $u'$ and $g'$, particularly near the northern and
southern survey limits at about 1.8 airmasses.

The images for the astrometric chips are almost as good as those over
the photometric array except for the 
outer half of field 18, the outermost of the first rank of CCDs, where
the images again reach two--thirds of an arcsecond in rms diameter.

The images for the focus chips (field 22) are still quite good, about 0.38
arcsecond rms diameter, and with almost no variation over the field of the
focus devices, so
even though the focus sensors are near the outer edge of the field,
the sensitivity of the focus servo is still essentially determined
by seeing.

\medskip
\noindent
{\bf 3. The CCDs and the Photometric System}\hfil

\smallskip
\noindent
{\bf The CCDs}\hfil

The whole SDSS project hinged on the availability of many (42, including
spares) 2048 $\times$ 2048 CCD sensors, for this camera, for a small
photometric monitor telescope, and for a pair of two--channel fiber
spectrographs.  At least 8 of these need excellent UV sensitivity and
very low readout noise ($< 5$ electrons.) At the inception of the work
more than seven years ago, it was by no means obvious that the chips could
be obtained.  The Tektronix 2048D, the 24 micron pixel 2048$\times$ 2048
device around which we were designing the optics (because it was a good
match to a reasonable--sized telescope with a reasonable focal ratio and
was the only commercial possibility at that time) was supposedly a
commercial item, but in fact was in very short supply, and without the
very large order for the devices from this project might well have been
discontinued.  Anxious moments continued to occur for a long time after
the order was placed and accepted, but the situation improved markedly
in the intervening years, and the requisite number of chips arrived,
worked, and are now operating in the camera.  We cannot thank the crew
at Tektronix/SITe enough for their patient persistence and unfailing
cooperation in this effort. 

Our cosmetic requirements were not as severe as those of many of the customers
of Tek/SITe, since TDI gets rid of a whole suite of defects
which would mar performance in normal imaging mode, and we have procured
chips with special grading qualitatively different from and in general
somewhat lower than that used for their normal Grade 1 devices. 

Our requirements differ quite widely from filter to filter, since in
$z'$ and $i'$ the expected sky levels are over 1000 electrons,
but in $u'$ they are only about 40 electrons, so quite noisy devices can
be tolerated in some places and very quiet ones are needed in others.  We
began by setting up a complex set of requirements, but in the end took
the results of the best efforts of SITe, which turned out to be quite
satisfactory.  The chips are also not all alike---we decided on economic
grounds to use frontside--illuminated devices for the $z'$ band, at what
we believed at the time to be a price of about a factor of two in
quantum efficiency.  This was a fortuitously exceedingly fortunate
choice which we have not regretted at all because of a later-discovered 
internal scattering phenomenon exhibited by thinned devices in the infrared
which would have made
data reduction with thin chips working as far into the infrared as $z'$
very difficult.  (We discuss this matter further below; it is not
a serious problem for our other bands, where we {\it do} use thinned
chips). For the detectors for $g'$, $r'$, and $i'$, we use
the standard visible antireflection coated (VIS/AR) thinned backside 
illuminated devices.
It also appeared at the time that special coatings for
the $u'$ were desirable to enhance the quantum efficiency in the
ultraviolet; the wisdom of this choice is now not so clear.  The
characteristics of the CCDs in the camera are summarized in Table 2. 
The mean quantum efficiencies are measured in 100 \AA{} wide bands at
the indicated wavelengths, and have a very small dispersion for all
chips except the $u'$ ones, where it is about 10\% of its typical value. 
The room temperature QE at 3500 \AA{} is between 40 and 50 percent with our UV
coating, but we have found that this high QE in UV declines at low
operating temperatures, and is between 35 and 40 percent for all our
devices at --80 C. The gain over the standard VIS/AR devices 
is not very large, though the steep falloff of their QE to the ultraviolet
would significantly lengthen the effective wavelength of the $u'$ band.
The expected sky fluxes are for the mean sky 
brightness at the site at 1.4 airmasses, scanning at sidereal rate.
The quantum efficiencies for thinned CCDs with the
normal and the UV--enhanced antireflection coating, as well as
unthinned devices, are shown in Figure 3.

The cosmetics and charge--transfer characteristics of the devices 
are in general very good. As part of the testing, CTE was measured at
a level of 200 electrons. It is typically better than 0.99999 both
horizontally (serial) and vertically (parallel). 
We measure CTEs 
of around 0.99998 horizontally and 0.99994
vertically at illumination levels of 30 electrons.
At the lower level, which is somewhat less than the
ultraviolet sky level, the net transfer efficiency is 86 percent from
the upper center of the chip, and 92 percent for the mean pixel in the
TDI scan. The effect on
the PSF is essentially to increase its RMS height by convolving it
with an exponential with an exponentiation length of about 0.4 pixel. The
seeing typically has a gaussian core with a sigma of about 1 pixel, so the 
core is widened by about 8 percent for objects at sky level, and less
for brighter ones; for objects at the detection limit the central pixel
is about twice as bright as the sky in $u'$, and the effect is roughly halved.
For PSF-fitting photometry, which we will use, the photometric error is
approximately half the width error for each dimension, so the error induced
is about 0.04 magnitudes at the sky, and about 0.02 magnitudes at the
detection limit in $u'$ (faint objects are biased slightly {\it bright}).
There is also a shift, of course; this is of
order of 0.02 pixel horizontally and 0.07
pixel vertically ( 8 and 28 milliarcseconds, respectively) 
after averaging over the column as TDI does. Again, these errors are much
reduced for detectable objects.

For the higher--background chips the astrometric shifts
are 0.007 pixel horizontally and 0.02 pixel vertically, and the photometric
errors completely negligible.
The overall cosmetic uniformity is excellent as well, with rms large--scale
QE variations of about 7\% in the blue and 4\% in the red and
infrared. The main defects we have seen are parallel traps of various
strengths, many of which are strong enough to cause serious CTE degradation
in the vertical direction with the backgrounds we are using, though, as
we discuss below, they are not fatal if they involve only isolated columns. 
A subimage in the corner farthest from the amplifier 
for a typical device operated with only one amplifier
of a Ronchi target with 30 electrons
signal in the bright bars is shown in Figure 4, in which the
excellent cosmetic appearance and low--level CTE are seen.

The full well is in the neighborhood of
300,000 electrons except for the $u'$ devices, where the parallel
gates have to be run at lower potentials in order to avoid shot noise
from spurious charge generation; we measure about 200,000 for them. The
signals in $u'$ are so much lower than in the other bands even for very
blue objects that this is not a problem. 

We specified in our selection criteria that no device can have adjacent
bad columns; this is driven by the fact that we are well enough sampled
that we can interpolate effectively over a single bad column using
linear predictive coding techniques, but cannot over two or more. The
number of single bad columns varies a great deal from device to device,
and goes from none to about 30 for the worst $z'$ chip. The frontside
$z'$ devices were produced early in the program and have the worst
cosmetics, but fortunately the signal levels are very high in this band.

These chips can in principle operate in multi--pinned phase mode,
but we cannot use it effectively in the scanning array, since it is
clocking all the time.  They are driven by three--phase clock signals
generated from high--stability DC rails with CMOS switches and all the 
chips are clocked synchronously.  The typical
output gain of the on--chip FET is 1 $\mu$V/electron, and the dynamical
range is 30,000:1 with full well of 300,000 electrons.  Dark current at
20\deg C is typically $<$200 pA/cm$^2$ for the front--illuminated and 
back--illuminated VIS/AR devices and somewhat higher for the UV devices;
in no case does the dark current produce appreciable signal or shot
noise at --80C.

The devices are designed and bonded out so that the four quadrants can be
read independently, which for many applications results in a factor of 4
improvement in readout time; for us, it is the split in the serial
direction which is useful, since we must integrate over the full chip
vertically. The requirement of reading out the whole device in 54 seconds
for sidereal rate scans results in a pixel time of about 24 $\mu$s with
the split serials. 
Unfortunately, this does not work for all chips in
hand: acquiring enough devices demanded that we accept a few (6) which have
only one good on--chip amplifier. For these, we must clock the serial
register twice as fast, which incurs a noise penalty of about a factor
of 1.3, so the one good amplifier has to be better. We generate
two synchronized serial clocking streams to service the one-- and two--amplifier
devices, one synchronized to but exactly twice as fast as the other.

The devices, first characterized roughly at SITe, were evaluated with a
cold test station at Princeton having electronics similar to the survey
electronics.  The system noise is about 1 electron, and we have the
capability to measure charge transfer efficiency (CTE) and QE as a
function of wavelength, uniformity, and also to test the vertical CTE in
TDI mode using a special parallel--bar target and a flash lamp.  The
technique involves running the chip for some time with a uniform ``sky''
background of appropriate level, and then exposing the bar target, which
consists of about 200 thin bright lines parallel to the rows of the CCD,
with a flash lamp to impose a low--level signal.  This frame of data is
then captured as the chip continues to scan in TDI mode, and the bars
are superposed to simulate a single such bar traveling along the chip
following the charge packets.  This test is important, because some very
low--level parallel traps which show up in single frames are satiated by
the sky in TDI mode and disappear; other stronger traps permanently
damage the CTE in the affected columns. 

We verified an effect first reported by Richard Reed (private communication)
which has a drastic
effect on the spatial resolution of thinned chips in the far red; there
appears to be a halo of the form $B \propto {\rm S~} \exp(-r/r_0)r^{-1} $, where
S is the point--source total signal, with a characteristic radius $r_0$
which is a strongly increasing function of wavelength; we found that
$r_0 \simeq 50\lambda^2$, with $\lambda$ in $\mu$m and $r_0$ in pixels, for
these devices.
The fraction of light $f$ in the halo is reasonably well represented by
the exponential relation $f=\exp(11.51(\lambda-1.05))$, essentially
unity for $\lambda > 1.05\mu m$. The fraction of light in the halo is
about 0.9\%, 5\%, and 30\% in the $r'$, $i'$, and $z'$ bands, respectively.
The phenomenon is apparently caused by the trapping
of transmitted radiation between the metallic solder surface used to
attach the translucent substrate of the thinned die to the package and
the silicon substrate. 
The surface brightness in this scattering
halo is roughly the same order as that from the atmospheric/optical
scattering wings for a small range of radii in $i'$, and slightly complicates
the algorithms we use to subtract bright stars, but is otherwise not
serious. Its effect is negligible in the shorter bands. 
If we had used thin chips for $z'$
we would have been in serious trouble, 
and would in fact have been no better off
purely from a signal--to--noise point of view than we are with the
supposedly less sensitive thick devices.  The signal from small sources
like stars is not very much greater in the thin chips than in the thick
ones with half the QE because much of it is scattered into the sky, but
the sky gets the full QE increase; the result is that the signal to
noise is essentially the same for faint sources with thick devices as
with thin, but the thick chips do not have the halos. 

We elected to take chips mounted in their standard kovar header packages
even though this led to significant mechanical difficulty in their
mounting and cooling; demanding better packaging would have precluded
culling our devices from a commercial production stream and would have
prohibitively increased the cost and probably made the endeavor economically
impossible.  The problems incurred are fairly serious, however.  The
expansion coefficient of kovar matches silicon (and the substrate for
the thinned devices, which does match silicon well) so poorly that the
overall curvature of the devices, already serious at room temperature
because of problems in high--temperature processing, is much worse at
operating temperature.  As mentioned above, the chips are convex toward
the incoming light by about 230 microns center to corner, and that value
roughly doubles in cooling to --80 C.  We have dealt with this problem by
cementing a heavy kovar stiffener, which is part of the CCD
ball--and--socket mounting which we discuss below, to the back.  It was
also necessary to build a precision measuring microscope to aid in the
gluing of the photometric chips to this mount, since we wished to
position the chips to subpixel accuracy and the chips are not mounted
very accurately in their headers.  We thus used reference points on the
die itself to reference the CCD to its mounting system, and succeeded in
doing so to an accuracy of about 3 microns rms. 

The frontside (thick) $2048 \times 400$ astrometric/focus CCDs were also 
produced by SITe for us in two foundry runs.
Except for the decrease in parallel gate capacitance because
of the decrease in the number of rows, the devices are electrically identical
to the photometric CCDs.
These chips have been mounted on precisely machined invar--36
headers of our design which allow them to be mounted quite close
together in the column (short) direction, as shown in Figure~1;
the minimum distance is in fact determined by the filters,
which must be oversized to 
allow for the f/5 beam.  The headers are machined with a slight
convex curvature to match the photometric chips in the long direction;
the chips are flexible enough that they bend to this curvature easily,
and are mounted to the headers with a thin thermally setting epoxy film.
The electrical signals to and from
the chips are carried by a kapton flexible printed circuit (FPC)
on each end with
very thin and narrow copper conductors.  These FPCs are mounted
permanently on the CCD headers and the chips are bonded out to pads
on them.

\smallskip
\noindent
{\bf The Photometric System}\hfil

Our photometric  system comprises five
color bands ($u'$, $g'$, $r'$, $i'$, and $z'$) that
divide the entire range from the atmospheric ultraviolet cutoff at
3000~\AA\ to the sensitivity limit of silicon CCDs at~11000~\AA\ into
five essentially non--overlapping pass bands.  The system was described
in detail in Fukugita \etal (1996) for the SDSS photometric monitor telescope,
which is identical to the system for the main camera except for the
fact that the monitor has a single UV--coated CCD for all bands and
slightly different mirror coatings. We review the system here and describe
in detail only those 
features which are unique to the camera or are associated with the
camera sensitivity.

The filters have the following properties:
$u'$ peaks at 3500~\AA\ with a full width at half--maximum of~600~\AA ,
$g'$ is a blue--green band
centered at~4800~\AA\ with a width of~1400~\AA , $r'$ is the red
pass band centered at 6250~\AA\ with a width~1400~\AA ,
$i'$ is a far red filter centered at~7700~\AA\ with a width of 1500~\AA ,
and $z'$ is a near infrared pass band centered at~9100~\AA\ with a width
of~1200~\AA ; the shape of the $z'$ response function at long wavelengths is
determined by the CCD sensitivity.
 
While the names of these bands are similar to those
of the Thuan and Gunn photometric system (Thuan \& Gunn~1976;
Schneider, Gunn, and Hoessel~1983),
the SDSS system is substantially different from the Thuan-Gunn bands.
The most salient feature of the SDSS photometric system
is the very wide bandpasses used, even significantly
wider than that of the standard Johnson--Morgan--Cousins system. 
These filters ensure high efficiency for faint object detection and essentially
cover the entire accessible optical wavelength range. 
 
The filter responses are in general determined by a sharp--cutoff
long--pass glass filter onto which is coated a shortpass interference
film, and thus exhibit
wide plateaus terminated with fairly sharp edges. The exceptions are the
$u'$ filter (the passband is defined by the glass on both sides and it is
much narrower than the others) and the
$z'$ filter (no long wavelength cutoff).
The division of the pass bands is
designed to exclude the strongest night--sky lines of O~I~$\lambda$5577
and Hg~I~$\lambda$5460.
The $u'$ band response is similar to TG $u$ and Str\"omgren $u$ in that
the bulk of the response is shortward of the Balmer
discontinuity; this produces a
much higher sensitivity (combining with $g'$)
to the magnitude of the Balmer jump at the cost of lower total throughput.
Proper consideration of photon noise indicates that this is to be
preferred to a wider band with dilution by redder light. The makeup
of the filters is detailed in Fukugita \etal (1996); the interference coatings 
were done by Asahi Spectral Optics Ltd., Tokyo.

The photometric filters are 57 mm square and are 
all brought to 5~mm thickness by adding a
neutral glass element (BK7 or quartz), so that all filters have
approximately the same optical thickness; this is necessary to keep the
scale variations on the CCDs small.  The neutral glass element (GG400
for $g'$) is a plano--concave lens with a radius of curvature of 670 to
770 mm, serving as a field deflattener as discussed above.  These
filters are cemented to the backside of the second corrector. 

The system response functions, $S_\nu$ are shown in Figure~5.  
The response curves
include the filter transmission, the quantum
efficiency for CCD, flux loss due to the correctors
and the reflectivities of the two aluminum surfaces.
Figure 5 also presents the response curves including atmospheric 
extinction at 1.4 airmasses, based on the
standard Palomar monochromatic extinction tables
scaled to the altitude of Apache Point Observatory (2800~m). 
The characteristics of the pass bands (with 1.4 airmasses)
are tabulated
in Table~3.

The response functions differ slightly from chip to chip. We define the
SDSS photometric system by the response function of the SDSS ``Monitor
Telescope", a 60~cm reflector located at the same site.
The telescope is equipped with a thinned, back--illuminated, 
uv--antireflection coated CCD, 
the same as those used for $u'$ imaging of the photometric array.
We will apply a color transformation to the instrumental magnitude
obtained with the photometric array, converting it into the system defined
by the Monitor Telescope detector. These corrections are small
(of order 0.02 magnitudes) except in the $z'$ band, which is significantly
redder and broader in the camera than in the monitor system owing to 
more extended infrared response in the thick devices and the steep wavelength
dependence of the scattering, which effectively depresses the quantum
efficiency for stars, in the thin chips.

In Table 3 
the quantity $q_t$ is the peak system quantum efficiency in
the system, and $Q=\int d(\ln\nu)S_\nu\approx q_t\Delta \lambda/\lambda$ 
is the system efficiency, which relates the monochromatic
flux averaged over the filter passband to the resulting signal
expressed in terms of the number of photoelectrons:
$$
N_{el} = 1.96\times 10^{11} t Q 10^{-0.4AB_\nu} \eqno{(2)},
$$
where $t$ is exposure time in seconds; a 27\% obscuration of incoming flux
by the secondary mirror and the light baffles are taken into account.
$AB_\nu$ is AB magnitude at frequency $\nu$, corrected for atmospheric
extinction.

Table 4 gives the expected sky background counts and expected 
saturation levels for the camera, assuming a zenith sky $V$
brightness of 21.7 mag (arcsec)$^{-2}$
and a Palomar sky spectrum (Turnrose, 1974) modified to remove the
strong mercury and sodium lines from Los Angeles and San Diego.
Saturation corresponds to an assumed full well of $3\times 10^5$ electrons.
The {\it effective sky brightness} in the tables is corrected for atmospheric
extinction so that it is treated properly along with the flux from the star.

The counts and signal--to--noise ratios for
for stellar objects with a
double Gaussian PSF with $\sigma=$0.38 and 1.09 arcsec
with the same sky are given in Table 5. 
The noise is assumed to be given by
$$N_{el}({\rm noise})=[N_{el}+n_{\rm eff}f_{\rm sky}+ 
 n_{\rm eff}{\rm RN}^2]^{1/2} \eqno{(3)}$$
for a noise--effective image size of $n_{\rm eff}$ pixels.
The exposure time is taken to be 54 seconds, and
the read noise RN is taken to be 7 e$^-$.  The dark current is
negligible.  

The limiting magnitude for detection, set at S/N= 5, will be
approximately $u'=22.1$ mag, $g'=23.2$ mag, $r'=23.1$ mag, $i'= 22.5$
mag and $z' = 20.8$ mag for stars.  Signal to noise ratio of 50:1 (and
hence photometry at the 2\% level) is reached at 19.1, 20.6, 20.4, 19.8
and 18.3 mag in the five bands.  Typical galaxy images reach a given S/N
half a magnitude to a magnitude brighter at the faint end, depending on
their surface brightness. 

\medskip
\noindent
{\bf 4. Mechanical Design of the Photometric Array}\hfil

The CCDs for the photometric array are housed in 6 long thin dewars
(Figures~6 and 7) machined from aluminum blocks, each containing 5
chips in one column.  The CCDs are kept cooled at --80\deg/C during
operation by an auto--fill liquid--nitrogen system which will be described
in the next section.  The optical system is so fast and the focal plane
so large that mounting the chips and maintaining dimensional stability
is potentially a difficult problem.  We require astrometry at the 30
milliarcsecond per coordinate level, which corresponds to 2 microns in a
focal plane 650 mm in diameter.  We have solved this problem in a rather
unusual but, we think, very satisfactory manner.  The final corrector in
the optical system is a quite thick piece of fused quartz with a flat
rear face, 45 mm thick in the center and some 8 mm thicker at its
thickest point; we use this element as the mechanical substrate to which
all the CCDs are registered and all the dewars attached.  The corrector
thus serves as both a mounting substrate and a window for the camera
dewars.  Quartz is strong, reasonably stiff, has excellent dimensional
stability and very small thermal expansion coefficient.  The small
mechanical deflections ($\approx 1$ micron) associated with loading it
with the camera have completely negligible impact on its optical
performance.  Figure~8 shows the front view of the whole assembly
through the final corrector as it would look with the shutters removed. 
Figure 9 shows the assembled corrector plate, mounted in its support,
with the dewar mounting rails. 

The CCDs in the column are mounted 65 mm center to center, which leaves
1.5 mm gaps between the 63.5 mm square Kovar packages in which the
devices are mounted.  There is a somewhat larger gap between the edges
of the packages and the side walls of the dewars.  It is important to
mount the CCDs in the camera so that each of them is adjustable for
rotation, tilt, and piston to high precision, and that this
precision is stably maintained for the whole period of the survey once
they are fixed. 

The CCDs are individually cemented to the inner cone element of a mount
which has double cone structure.  The outer cone, made of invar--36, acts
as a socket. It has machined into its base a precise 45--degree cone which
engages a spherical surface on the inner cone. This piece, made of Kovar, 
is movable within the
socket and works like a ball floating in the outer socket within a
restricted angle range, with the center of curvature at the surface of
the chip (Figure 10; see also Figures 11 and 12). The surface of the ball
is mostly cut away, so that three small segments of the sphere 120 degrees
apart engage the conical socket. This ensures that the inner ball part
sits stably in the outer part even if the conical socket is somewhat
distorted. These ball--and--socket mounts provide
tilt adjustment via a set of four push--push screws diagonal to the chip
direction close to the edge of the cone on each socket (see the leftmost
cone in Figure 12 where screws are seen).  These assemblies are in turn
mounted on a `T'--shaped super--invar optical bench (the ``Tbar'')
in a manner which allows small independent
rotation of the chips and shimming for piston.  The Tbars are stiff
laterally, reasonably stiff vertically but quite limber in torsion.  The load
of their own mass and that of the CCDs with their double cone mounts
amount to about 2.7 kg.  Under the full gravitational load the worst--case
deflections are 2 microns in the focus direction and 1 micron in the
image plane when moving from the horizon to the zenith, somewhat less
over the operating range of the survey (above 35 degrees elevation.)

Two important desiderata for the mounting of the Tbar on the quartz
corrector are thermal isolation and mechanical stability against
deflection.  These requirements are fulfilled by mounting the optical
benches, one per CCD column, by a ball--and--rod kinematic mount, which
consists of a quartz column bonded and screwed to the corrector, and a
set of four ball--and--double--rod pads (Figures 11, and 13a,b for an
enlarged view), one assembly at each corner of the Tbar.  
On one end of the Tbar, the arrangement consists of two balls on
sets of parallel ways, one parallel to the long axis of the bench and
the other perpendicular, which locates that end in both dimensions to
within small rotations (Figure 13).  On the other end, one ball rests in a
set of parallel ways parallel to the bench, which with the set at the other
end fixes the rotation but
is free to move along the bench. The ball rests in sets of {\it mutually
perpendicular} ways, which is completely free to {\it slide} in the
plane.  With this structure the effect of different thermal expansions
of quartz and invar are buffered, and, more importantly, manufacturing
variations do not affect the kinematic nature of the mount. 

The bench is sufficiently flexible in torsion that the four vertical
constraints can be mated independently with quite reasonable dimensional
tolerances and forces (50N) on the balls, and in fact needs the
four--point support for torsional stability.  The $1/4''$ balls are made
of titanium, which is tough, and combines reasonably good Young's modulus
and reasonably low thermal conductivity. The $3/32''$ rods are of hard
302 stainless steel.  The conductive losses for each ball joint are
about 0.5 watt, with the 100C temperature drop shared roughly
equally between the ball joint and the quartz column.  The heat flux
across the mount can be calculated to be $(\Delta T*{\rm conductivity})*
({\rm force*radius/elastic~modulus})^{1/3}$. The measured conductivity
of the ball--rod joint was within about twenty percent of the calculated
value for a wide variety of materials we tried.  The scheme appears to
work very well.  We expect a total of about 2 watts of heat loss through 
the kinematic
mount.  The total deflection in the ball--rod mounts under their static
50N load is 3 microns; in addition each ball can support as much as 7N
vertically and 20N longitudinally from variable gravitational loading. 
The resulting differential deflections from the zenith to the survey
altitude limit is of the order of 0.8 microns. 

The one disadvantage we know about this scheme is that the stresses on
the balls and rods are very high owing to the tiny contact area.  There
is no danger of failure at the static stress levels, but dynamic loading
associated with handling could easily permanently deform either member. 
To avoid this, each dewar uses a set of four electroformed nickel
bellows which are pressurized with dry nitrogen at about 300 kPa (these
exert the roughly 10N static forces referred to above) to bring the
optical benches into contact with the kinematic mounts for observing
(Figure 14).  When the camera is being mounted or moved, this pressure
is relieved and springs retract the optical benches about 1 mm. At the
same time the retraction of the bellows engages a latch at each end of
the Tbar which snaps into place as the Tbar retracts and latches it 
away from the kinematic mounts. 

In general, since the loading changes as the survey goes on are very slow
and the astrometric calibration through astrometric standards is a
continuous process, the contribution to the astrometric error from the
deflections discussed above should be negligible; in any case, the
errors are much smaller than those expected from telescope deflections
and drive irregularities.  The overall deflection of the corrector in
the focus direction (which is the {\it only} direction there are
appreciable deflections) is about 2 microns neglecting the stiffening by
the dewar bodies.  At worst, focus changes induce centroid motion (from
an overall scale change) a factor of 20 smaller, so there is no
appreciable error from this source.  Since the dewars in this design are
simply vacuum enclosures and, except for the flexibly coupled preload
bellows pushrods and thermal straps (see Section 6), do not even contact
the optical benches, the load paths are very direct from the kinematic
mounts to the telescope structure.  The only tricky part of the design
is the fact that there was a fair amount of machining to do on the
quartz corrector.  There are about 100 holes for screw anchors for the
kinematic columns and the dewar mounting rails.  The screw anchors
consist of knurled brass inserts epoxied into these holes (see Figure 9).
The whole
process went without mishap, though there was a fair amount of anxiety
in the beginning, as there was in applying the striped coatings onto the
corrector. 

Thermal stability is a major concern for the Tbars as are
the physical dimensions; locating the CCDs in all three dimensions to
the required accuracy was not easy. The photometric Tbars are made of 
superinvar (Fe--32Ni--5Co),
whose thermal expansion coefficient is around $0.4\times 10^{-6}$ at room
temperature. The Tbar, because of its complex shape and the required precision,
had to go through a special manufacturing process.
The machining was done in the new shops at the National Observatory
of Japan in Mitaka with extensive use of wire EDM in order
to minimize mechanical distortion. The Tbars were annealed
(600C for 1 hour) five times, once after every major cutting
operation, in order to release mechanical stress and ensure
stability. The last annealing was done by holding the temperature at --100C 
(near the detector operating temperature) for 1 hour and then raising it to 
200C for 1 hour.
The final measurements of the mounting face of the Tbar show that deviations 
from an ideal flat plane were typically smaller than 20 microns, and most
of this was essentially pure twist. The 50N forces holding the Tbar to the
kinematic mounts are more than sufficient to flatten this.
 
It is necessary to adjust the tilt, rotation, and focus of each CCD on the
Tbar
fairly exquisitely; we allow 25 microns tilt error, 5 microns rotation
error, and 25 microns total piston error.  The tolerances on the
absolute x,y location of the CCD are not so severe, though we succeeded
in locating all the chips within a about a pixel (24 $\mu$m).

The tilt adjustment was done on initial assembly of the ball and socket
joint to theoretical optical design values.  It can be changed on the
basis of later tests but only with some difficulty; we feel, however,
that it is the most reliable of the adjustments to predict.  
The rotation and tilt adjustments were done on a
measuring machine which incorporates an accurate XYZ stage carrying a
microscope and a precision linear slide upon which the Tbar is mounted on 
kinematic
mounts like the ones on the corrector (Figure 15). Precisely
made stepping 
blocks allow positioning the Tbar on its slide so that each CCD in
turn is positioned accurately at the same position in front of the
microscope stage by the insertion of one more block.  
The microscope has a very fast inverse--Cassegrain
objective which allows z--motions to be measured using focus alone to an
accuracy of about 2 microns; a crosswire reticle allows positioning in
the image plane to about 1 micron.  The three--dimensional coordinates of
six reference points on the CCD die (four at the corners and two at the 
serial register
splits) are obtained and the adjusting screws manipulated until the tilt
and rotation are correct; the rotation is adjusted by means of a long
(250mm) rigid lever attached temporarily to the socket part of the CCD mount and
moved with a micrometer.  Though the adjustments are not in themselves
micrometric and a certain amount of iteration was necessary the
procedure converged reasonably quickly; the adjustment of a Tbar was
typically accomplished in less than one day.  Piston was set with a set
of shims to match the design focal plane (see Figure 11 above) and it
may be necessary to refine these (and the rotation) using measurements of the
sky, though the measured parameters of the as--built optics indicate
that the real focal plane will deviate from the design one by no more than
10 microns.

The measurements show this procedure is quite accurate; the rotation is
fixed within 3 microns RMS, and the error of tilt and piston is together
only 10 microns RMS, which results in image degradation of only 2
microns and tiny (less than 1 mas) maximum astrometric errors. 

The dewar bodies themselves are machined of aluminum (see Figure 7), and
have O--ring seals to the quartz in front and to a fitted lid which
carries the cooling system and electronics in back (Figures 16 and 17). 
They are about 75 mm tall and 330 mm long, so the atmospheric pressure
on the side walls results in about 2500 N when they are evacuated.  This force
is taken up by a lip on the lid of the dewar in back and in front by a
frame machined integrally into the piece, which consists of horizontal
stiffening bars between the filters.  Thus the forces do not act on any
dimensionally critical element.  A similar force in the focus direction
acts to seal the quartz to the dewar body, and demands that the face of
dewar body be quite accurately flat in order that it not distort the
rear surface of the corrector.  This would have no optical consequence,
but would change the effective shape of the focal plane.  This strong
bond with the dewars results in considerable stiffening of the corrector
by the dewars, since the Young's moduli of aluminum alloys and quartz
are similar; though the dewar walls are relatively thin (about 1 cm)
they are twice as deep as the corrector is thick, and most of the
stiffness is, in fact, in the dewars.  The great disparity of their
thermal expansion coefficients, however, demands that the joint have
rather low friction; thin polyimide gaskets appear adequate to the task. 
When the dewar is removed, cleats machined into it bring the optical
bench away with it for ease of maintenance.  The optical bench can then
be removed from the dewar through the top, carrying the in--dewar circuit
boards with it after they are detached from the wall of the dewar and
attached via special fixtures onto the optical bench. 

\medskip
\noindent
{\bf 5. The Astrometric/Focus Array}\hfil

\smallskip
\noindent
{\bf Concept}\hfil

We need reasonably good relative astrometry to place the fibers for
spectroscopy.  Allowing for errors arising from differential refraction
with wavelength and position, hole placement uncertainties, etc, it
seems necessary to find positions to accuracies of the order of
$\approx$200 mas (12 microns in the focal plane).  This is difficult to
do with the photometric array, because it saturates at about fourteenth
magnitude in the bands most useful for astrometry ($g'$ and $r'$, with
$r'$ preferred because of the smaller refraction corrections), and there
are very few astrometric standards at this brightness level.  It would
be in principle possible to calibrate the camera astrometrically, but
one would have to depend on its stability and the stability of the
telescope drives for relatively long periods to make use of the
calibrations without suffering intolerable overheads.  This problem can
be solved, however, by using the astrometric array of CCDs, which allow
us to tie much brighter astrometric standards to the postions obtained
from the photometric array. 

The astrometric/focus camera uses 24 $2048 \times 400$ frontside 
CCD chips working
in the $r'$ band and mounted in the available space above and below the
photometric array (see Figure 1).  With these chips the integration time
is reduced by about a factor of 5 (400/2048) relative to the 54 second
photometric exposure time; the chips are front illuminated devices
without coatings, which are less sensitive in the red than the thinned
photometric chips by a factor of about two, so without further
filtration the saturation level is brightened by about 2.5 magnitudes,
to about 11.3.  This is still not particularly useful, since the
Hipparcos/Tycho net and the AGK3 (whose positions are probably not good
enough anyway) both have very few objects as faint as 11.3.  To reach
brighter standards, we employ neutral--density filters with 3.0
magnitudes of attenuation, which, together with shorter integration
times and the use of front--illuminated devices, enable us to observe
astrometric standards as bright as $r'$=8.3. 

The disadvantage of the shorter columns is that the shorter integration
times lead to larger position errors because of seeing--related image
wander.  With an 11 second integration time in one--arcsecond seeing,
Kolmogorov seeing theory suggests that the one--dimensional positional
accuracy achievable in that time is about 50 mas with the two passes for
each star.  We will see below that we encounter about one astrometric
standard every 10 seconds or so, so the ultimate astrometric accuracy is
critically dependent upon the short-timescale accuracy of the drives (as
well, of course, on the short-timescale dimensional stability of the
camera and telescope, but we expect negligible errors from these sources). 
Given sufficient stability so that we need to make negligible corrections
on the one-minute effective exposure timescale, we should be able to
{\it standardize} to of the order of $50/\sqrt 5 \simeq 20$ mas.
 
The response of the astrometric detector system is given in Figure 18,
where the sensitivity of the astrometric system has been multiplied
by a factor of 10 to facilitate comparison with the photometric system.  
The filters, 15mm wide and 55 mm long, 
are composed of a 2.5mm thick piece of NG4 and a 3.3mm thick
piece of OG550 with the short--pass interference coating on the free
surface. The OG550 is figured on the coated surface to act as a field
deflattener; the astrometric CCDs are manufactured to have the same 
curvature as the mean photometric device. We
intended the filters to have the same pass band as the photometric
$r'$, but the short--pass interference coating on them, which was
supplied by a different vendor from the $r'$ filters in the photometric
array, was delivered with a cutoff wavelength 50 nm too long.  It was
deemed in the light of schedule pressure that replacement was not
warranted; the main effect is that the attenuation is slightly less than
desired. 

These chips can be run somewhat warmer than the photometric imaging
devices because we are not interested in low signal--to--noise objects and
the integration time is short.  Cooling to --60 C is sufficient,
which yields a background of about 10--20 electrons per pixel in the 11
second integration time.  This background is actually useful; the
astrometric chips display rather poorer horizontal CTE than the
photometrics for reasons which may be connected with crosstalk in the
clock signals in the FPC connecting the chips to the electronics, and
the effect is lessened in the presence of even such a small background.

The centroid of a star which deposits about 2000 electrons ($r'\simeq
16.6$) can be measured to about 30 mas (shot and readout noise errors
alone) in 1 arcsec seeing with this background.  A star which saturates
in the central pixel ($4 \times 10^5 e^-$ for these devices, $r' \simeq
8.3$) has a total signal of about $4\times 10^6$ electrons, so the
dynamic range for 30 mas accuracy, considering shot and read noise
alone, is about 8.3 magnitudes.  This yields an overlap of about 2.5
magnitudes between saturation of the imaging array (about 14.0 mag) and
the 30 mas accuracy limit for the astrometric array. There are more than
200 stars per square degree in that magnitude interval near the pole, so
that of order 10 are on any photometric CCD at any given time.  Thus the
frame defined by the astrometric chips and the one defined by the red
imaging chips (and with only a little more difficulty, that defined by
any band in the imaging array) can be tied together very accurately. 

The array uses 12 devices in each of two nearly identical dewars, one
leading the photometric array and one trailing; one device in each dewar
(which has a special stepped--thickness $r'$ filter but no
neutral--density filter) is devoted to monitoring the camera focus.  The
sensors at the trailing edge provide a check on the tracking rate and
direction as well as an independent astrometric calibration; stars cross
the bottom set 7.5 minutes after they cross the top, and they go through
the red photometric sensors 1.4 minutes after they leave the top.  Thus
drive errors with frequencies lower than $2 \times 10^{-3}$~Hz can be
corrected for, and we can monitor higher frequencies statistically.  The
astrometric and red sensors define a continuous frame which will drift
slowly with respect to any initially defined absolute frame because we
are basically measuring rates.  We will use the astrometric standards
only to pin this instrumental frame to the sky at intervals. 

There are about 11 Hipparcos/Tycho stars per square degree at the pole in our
magnitude interval, and the CCD array will encounter one every 10 seconds on
average.  If the drives can be held to the accuracy we would like (25
mas rms stochastic component in the frequency interval $2\times 10^{-3}$
to $3\times 10^{-1}$ Hz), as seems very likely from measured errors on
the manufactured parts, we should be able to tie the survey to the
Hipparcos net to an accuracy of better than 50 mas.  In order to measure
positions well enough to place fibers for the spectroscopic part of the
survey it is necessary to have errors of the order of 200 mas or less,
and the seeing places limits of 30 or 40. Various scientific desiderata
demand, of course, arbitrarily high precision. Bettering the limits
imposed by seeing would have been prohibitively difficult and expensive,
but we decided early in the project to attempt to keep the errors near
that limit. This has been relatively easy and inexpensive to do in the 
design and construction of the camera; the situation for the telescope
and its bearings and drives is not at this writing quite so clear.

\smallskip
\noindent
{\bf Mechanical Design of the Astrometric/Focus Array}\hfil

The linear dimension associated with the expected 30 mas seeing error is
about 2 microns, and achieving dimensional stability to this order over
a focal plane as large as ours is not an easy task.  Invar--36 has a
thermal coefficient of expansion of about $1.5\times 10^{-6}$/\deg C,
and a one--degree temperature change induces a dimensional change of
0.7$\mu$ over the 455 mm width of the array, so one--degree temperature
control is adequate.  The coefficient for silicon is similar, and
changes on the scale of one chip with reasonable temperature control are
negligible.  Our approach to the astrometric array is thus very similar
to that for the photometric except for the mounting of the CCDs; the
invar--36 headers for the astrometric chips are mounted with screws onto
an invar--36 optical bench, which goes across the array, on invar shims
machined with the tilts necessary to fit the field curvature.  The
astrometric Tbar is shown in Figures 19 and 20.  The shorter vertical dimension
means that rotation is not so critical, and shimming for piston will be
the only adjustment normally performed.  The optical bench is located to
the corrector with kinematic mounts of the same sort as used in the
photometric dewars, but here the pillars are much smaller, shorter, and
are made of invar instead of quartz (See Figure 9). The optical bench
is housed in an an aluminum dewar attached to the corrector in the same
fashion as the photometric ones, again with an O--ring seal against the
quartz.  Reinforcing the dewar against transverse atmospheric pressure
is trickier in this case than in the photometric case because there is
no clear place to put a spreader bar close to the focal plane with the
chips overlapping as they do, but spreaders behind the main horizontal
web of the optical bench through holes in the vertical web serve here. 
This complicates the assembly somewhat, but not overly.  A dewar of
either variety can be disassembled to the point of removing the optical
bench or reassembled in of the order of an hour. 

The dimensional relationship between front and rear ranks of the
astrometrics is also crucial to control, since it determines the
accuracy with which errors in the drive rate and direction can be
measured.  This is greatly facilitated by our use of the quartz
corrector as the metering substrate.  The dimensions should not be much
less accurately maintained than those within one of the optical benches,
particularly when one considers that the deflections associated with the
kinematic mounts are the same for the front and rear benches (and
furthermore quite similar to those for the photometric benches). 

\smallskip
\noindent
{\bf The Focus Chips and Focus Servo}\hfil

One of the major contributions to image degradation in normal observing
circumstances at most telescopes is the inability to keep up with focus
changes brought about by flexure and temperature changes.  Our
requirements are especially severe because of astrometry, but it appears
that a simple scheme will suffice to provide excellent control. 

There are two astrometric--type sensors housed in each of the
astrometric dewars
which are used as focus sensors; though they are quite far from the
center, the image quality is sufficiently good at their locations that
they are ideally suited to monitor the focus.  They have only an $r'$
filter with no neutral density; this filter is cut into three parts, and
the optical thickness associated with the neutral filter on the sensors
is taken up by three clear glass (BK7) spacers of varying thickness (1.9 mm,
2.8 mm and 3.7 mm in addition to a 3.4 mm Schott OG 550 glass element). 
When the rest of the array is in focus, the center of the focus chip is
also, but the ends are, respectively, 300 microns inside and outside of
focus; this defocus, which results in image degradation from defocus
comparable to the expected 1 arcsecond seeing, is optimal from a
focus--determination signal--to--noise point of view.  In this way
comparison of the images in the two outer thirds will provide a
sensitive differential measure of the focus, which we will adjust
dynamically; the resolution of the secondary motion is such that it
should cause us no difficulty.  Figure 21 shows images at five positions
on the focus sensor through focus with 150 micron focus steps, as
produced by the design optical system and convolved with 0.8--arcsec
Gaussian seeing. 

We need to control the focus very accurately to maintain astrometric
accuracy.  If we require 2$\mu$m positional accuracy in the focal plane,
we must control the focus to about 35$\mu$m, since the maximum angle
which the central ray makes to the focal plane is about 0.055 radian. 
Focus errors of this size contribute only 7$\mu$ rms to the image
diameters and are negligible, but we would like the focus errors to be
negligible for the astrometric determinations.  The factors which
contribute to focus errors are residual aberrations across the field of
the focus chips, photon statistics, and, most important, seeing.  The
focus errors from aberrations and photon statistics (the actual errors
will almost certainly be dominated by seeing) are about 3 microns for
any star brighter than about $r'$=15.5, the limit set by PSF variations
across the focus chip.  For fainter objects, photon noise becomes
important and the errors are about 6 microns at 17 mag, 9 at 18 mag, and
20 at 19 mag.  At the galactic pole, the star counts are about 160, 350,
550, and 880 per square degree brighter than 15.5, 17, 18, and 19 mag,
respectively, and fainter than the saturation limit at $r'$=11.3 mag. 
The three focus zones are each about 4 arcmin wide, so the camera will
see about 0.0167 square degrees per minute of time in each of the zones. 
Thus it will see 3 stars brighter than 15.5 mag, 6 brighter than 17 mag,
9 brighter than 18 mag, and 15 brighter than 19 mag per minute.  The
timescale for focus changes is not well known yet, but the thermal
time constant for the telescope is of the order of an hour, so it seems
not unreasonable to expect that one can effectively average over a
hundred to a few hundred stars to generate a focus signal, and the
statistical errors are expected to be negligible. 

\medskip
\noindent
{\bf 6. The CCD Cooling System}\hfil

The dark currents for the Tek/SITe CCDs range from 30 to about 200
pA/cm${^2}$ at 20C.  At -80C, this is reduced by about a
factor of $3\times10^{5}$, which for a 24$\mu$ pixel becomes 0.01
electrons/second for 100 pA at room temperature.  The sidereal--rate
exposure time of 54 seconds thus yields a dark signal of about 0.5
electrons, which does not contribute significantly to the read noise. 
We could have chosen to run the devices warmer without serious
degradation of the data, though considerably more care in monitoring
the dark current contribution to the background would have been necessary.
The thermal losses at -80 C are roughly evenly divided between conductive
(which scale linearly with the difference between operating and room
temperature) and radiative (which are almost independent of operating
temperature), and operating warmer would not have made a major impact
on the thermal design. The CTE of the detectors, which is not outstandingly
good by today's standards, seems to be influenced not at all by temperature
over any interesting range. It is true that we could have bought some
extra sensitivity in the $z'$ chips by operating warmer, but they, the
oldest and hence earliest-produced devices in the camera, have cosmetics
which are much better at low temperatures, and the extra sensitivity is
mostly bought by extending the response into the infrared where the optical
performance is not very good. The savings in liquid nitrogen consumption
which might be realized by operating warmer are also not very great. All
these factors and a general conservatism (``We'll design it to run cold, but
if we have to back off it will not kill us'') led to the choice of
operating temperature.
 
The total thermal losses for the dewar design we are using come to about
10 watts per dewar, three watts from radiation from the detectors
(mostly from the active detector area itself), 2 watts through the
kinematic mount, 1 from radiation to the optical bench (which has a
gold--plated thermal shield), 1 from radiation to the nitrogen container
(also gold--plated with a secondary shield), about 2 watts from the
temperature--control makeup heaters, and another 1 from miscellaneous
sources such as conductive losses through the flexible printed circuits
(FPCs) connecting the CCDs to their support circuitry, conductive losses
to the force actuators on the kinematic mounts, etc. 

We use an auto--fill LN2 cooling system similar to the one used on the
4--shooter camera at Palomar (Gunn \etal 1987).  Each dewar 
has a small LN2 container holding about 400 ml of liquid (Figure 22), 
which keeps
the detectors cool for a little less than two hours.  Under normal
operating conditions all the LN2 dewars vent through a common
vacuum--jacketed fill line and are kept under moderate pressure.  There
is a temperature sensor on the LN2 dewar body which is monitored
continuously, and the dewar is filled when its temperature rises, but
normally the dewars are filled on a schedule kept by the executive
microprocessor (see below---we currently fill once an hour) which ensures
that they never go empty.  The fill is accomplished by opening all the
solenoid valves on the individual vent lines and allowing liquid under
pressure to enter; these are closed one by one as the dewars fill (as
indicated by thermistor--based liquid sensors in the vent lines).

The heat is conducted from the chips through the kovar cone mounts
cemented to them, and thence to copper posts which are connected via
multilayer silver straps the individual leaves of which are 50 $\mu$m
thick (silver is only slightly more conductive than copper but is nearly
a factor of two more flexible at a given cross--section) to a cold--finger
on the base of the LN2 container, which houses a Zeolite molecular sieve
getter.  The very low heat conductivity of invar makes it difficult to
keep the optical bench isothermal, but we keep the chips at a uniform
constant temperature with small make--up heaters associated with each CCD
and mounted to the copper posts, as seen in Figure 23, which attach
directly to the CCD mounts.  It is possible because of the low
conductivity of invar and the rather poor thermal contact between the
CCDs and the optical bench to operate the devices at quite different
temperatures if desirable, and we may choose to operate the $u'$ devices
colder to minimize their somewhat higher dark currents, though it is not a
serious problem at --80 C.  The thermal expansion coefficient of invar is
so small that small deviations from isothermality have little effect on
the dimensional relationship in the bench.  The temperatures are
measured with small platinum resistors fed with resistive dividers which
result in a reasonably accurately linear response over the relevant
temperature range.  Sensors of the same type are used to monitor the
temperature of the LN2 container. 

The situation for the astrometric dewars is quite similar.
At an operating temperature of --60 C the total heat load to the
astrometric optical bench is about 6 watts, roughly the same as the
photometric dewars.  We have decided for simplicity and economy to use
a cooling system essentially identical to the one adopted for the 
photometric dewars, and in fact one which shares many parts with its
photometric counterpart. For astrometric accuracy, 
we must endeavor to keep the bench as isothermal as
possible.  The bench has a cross--section of about 7 cm$^2$, and the
conductivity of invar is about 0.2 watt/C--cm, so a flux of 1 watt
in the bench is associated with a temperature gradient of about 0.7 C
per cm.  Temperature differences across the bench of 1 C
induce a bow in the bench with an amplitude of about 1 $\mu$m.  We 
distribute the heat load from the bench to a heavy cold
bar attached to the LN2 container's cold--finger at four points along
the optical bench, each point associated with a makeup heater and a
temperature sensor. The connection is again made using thin silver straps. 
This results in temperature inhomogeneities
of the order of 1 degree, but in a pattern which should remain quite
stable and hence astrometrically innocuous.

On the saddle which carries the power supplies (see Section 8) are mounted
two 10--liter intermediate supply dewars which supply the auto--fill system,
each one supplying LN2 for one astrometric and three photometric dewars
through a vacuum--jacketed transfer line into the camera body and thence
along thinwall stainless tubes insulated with closed cell polyethylene
foam to the individual CCD dewars. 
These in turn are kept filled from a 160--liter dewar on the rotating
floor of the enclosure, using an auto--fill system of necessarily 
rather different design but similar principle.

%
The approximately 10 watts of dissipation in a dewar means that about
250 ml of LN2 is consumed per dewar every hour.  Including 6 dewars and
two astrometric dewars, the total consumption of LN2 is about 2 liters
per hour, or 50 liters per day.  Thermal losses inherent in venting the
camera dewars back through their intermediate supply dewars increase
this consumption by about a factor of 1.5, and losses incurred in
transfer roughly double {\it this} with the system in its present
preliminary state.  We hope to reduce the consumption significantly when 
the system is properly tuned. 

\medskip
\noindent
{\bf 7. The Electronics}\hfil

\smallskip
\noindent
{\bf Overall design} \hfil

The electronic circuitry for the camera is implemented on 207 circuit
boards of 31 basic types. The boards are a mixture of thru--hole and
surface mount PC cards, and in addition there are 
several one--of--a--kind logic and utility boards which are 
wire--wrapped. For the most part, the boards can be accessed and replaced
without serious disassembly (not true of a few boards in the dewar vacuum)
and it is planned to keep a complete set of spares with the camera.
The object of the design was to make the system as modular as possible,
and to eliminate as far as possible hand wiring of boards, cables, and
connectors. In this we feel that we were moderately successful.
We discuss here primarily the active boards associated with the detectors,
and describe briefly some of the other circuitry.

The conceptual diagram of the CCD electronics is shown in Figure 24.
As in any CCD system, the electronics consists basically of two streams,
the digital signals which control and clock the CCD and the outcoming
analog video signals which must be processed and digitized.  The clock
waveforms are generated by one of two on--board microprocessors, the
{\it phase micro} which forms the intelligence for
the {\it CCD controller}.  These signals are bussed to all the
dewars over an RS485 bus and are received and converted to CMOS digital
signals by the {\it bus receiver board} in each dewar.  The rail
voltages are generated by digital--to--analog converters (DACs)
and buffered on the {\it bias board} for
each chip.  The CMOS clock signals and the rail voltages go into the
vacuum to a small {it clock driver} board, where they are combined to
generate the clock signals at the proper levels to drive the CCD.  The
video signals from the output FETs on the CCD go to a small {\it preamp}
which is mounted piggyback on the clock driver board, one per detector. 
There they are amplified by a factor of 10 and converted to a
low--impedance analog signal.  It emerges from the vacuum and goes to a
{\it signal chain board}, which is physically the other side of the bias
board, where the signal is level shifted, double--correlated sampled and
digitized.  The serial output stream from the analog--to--digital converters
(ADCs) on the signal chain
board goes to the {\it serial shift register board}, one per dewar,
which formats the digital output from single--amplifier chips into two
data streams with the same timing as those from a two--amplifier chip. The
ten resulting serial streams then go to the {\it FOXI transmitter} where
the streams are combined, converted to parallel, and then back to serial
at 100MHz to a fiber driver which sends the data to the data acquisition
system in the operations building. The controller also handles various
houskeeping chores, such as setting the bias DACs, monitoring the resulting
voltages, controlling the shutters, monitoring various system pressures
and temperatures, etc, and generating and driving the RS485 bus signals
and master 8MHz clock signals. Most of the control jobs are handled by the 
{\it executive micro}, whose RS232 line is one of two control links to
the outside world.
The numbers of the boards of various types and their electrical power
dissipation is given in Table 6. 

The clock drivers and the preamps are mounted in the dewars to minimize noise,
ensure ground integrity, and minimize
danger from electrostatic discharge,
as was done in the Palomar systems (Gunn and Westphal 1981).
The optical design has sufficiently good distortion
characteristics, as we discussed in section 2, to allow clocking
all the chips in the array synchronously, and hence the driver and digital
control circuitry is relatively simple.  
The preamp, clock driver, and signal chain/bias boards
each have two independent channels corresponding to the split serial registers
CCDs. Each CCD is associated with either one or two 
analog signal chain channels, depending on whether it has one or two good
amplifiers on the serial register in use. One channel on the boards belonging
to single--amplifier chips is idle. Most dewars have only one single--amplifier
device, but one has none and another has two. This peculiar circumstance
was demanded by the distribution of readout noise and the acceptable
noise thresholds for our five bands.

\smallskip
\noindent
{\bf The CCDs: Electrical characteristics} \hfil

The CCDs electrically are typical three phase devices. The parallel
waveforms as generated by the clock driver are sketched in Figure 25; 
T is the transfer gate signal. P3 is an MPP gate and must be clocked
about 3 volts more positive on its positive rail than the other two
in order to buck the implanted potential for normal operation. We clock
all the time, so there is not really an `integration' phase, but we hold
P1 and P2 both high during the serial transfer time, which is the bulk 
of the time.
The serial transfer signal (S1, S2
S3) is sketched in Figure 26. SW and R are signals for the summing well and
reset, respectively.  There are four other signals (IPC, I$+$, I$-$, and IR)
synchronous with these
for controlling the input clamp on the preamp and the double--correlated
sampling integrator on the signal chain board. Recall that there are two
sets of these serial signals a factor of two apart in frequency for the one--
and two--amplifier chips.

The architecture of the serial registers on the devices is such that
there are 20 `extended' pixels between the edge of the imaging array and
the on--chip source follower amplifier, and we will read another 20
overscan pixels in the two--amplifier chips. Hence each half--row will
consist of 1024 + 20 + 20 = 1064 pixels.  The single--amplifier chips
will have the 20 leading and trailing extended pixels at either end of
the data pixels and will in addition have 40 overscan pixels at the end. 
At sidereal scanning rate, the scale of 3.643mm/arcminute corresponds to
38.05 lines/civil sec, or 26.28 ms/line.  The vertical transfers require about
900$\mu$s, leaving 25.38 ms per line.  We use 30 master clock cycles
per pixel for all the functions in a classical dual--slope
double--correlated sampling system (see Figure 26), 
so a 1.3333 MHz (750 ns) clock yields
22.5--microsecond pixels.  (The fast channels run at 11.25$\mu$s per
pixel.) At 22.5 $\mu$s per pixel, the time for 1064 pixels is 23.94 ms,
safely less than the 25.38 we have for this activity scanning at the
sidereal rate. The 750 ns pixel clock for the normal chips and 375 ns
one for the fast chips is  derived from an 8 MHz master clock,
which is needed by the A/D converters and the data transmitters.

The photometric CCDs are electrically connected to the
preamp--clockdriver assembly via a 25--conductor kapton flexible printed
circuits (FPCs) with very thin copper conductors (see Figure 27).  
This FPC terminates
in a connector on the preamp board and is soldered to a BT--resin PC
card which acts as a socket and fanout board for the CCD on the other
end.  The astrometric chips, as discussed above, have no pins but are
built with two FPCs as part of their structure. We have used BT--resin
and polyimide boards throughout in the vacuum to avoid bromine contamination
from the flame--proofing in ordinary FR4 (``G10'') fiberglass boards. This
is standard practice in the high-energy physics community; we have no
direct evidence of corrosive damage from vapor-phase bromine, but the dewars
have a lot of electronics in a relatively small vacuum volume, and we 
chose to be cautious.

\smallskip
\noindent
{\bf Preamps} \hfil

The photometric preamps and the clock drivers are two small 
(43 mm $\times$ 53mm)
piggy--back BT--resin printed circuit cards using mixed 
through--hole (large hermetic tantalum electrolytics and
the preamp operational amplifiers) and surface mount (everything else)
technologies mounted on the inner wall of
the dewar (Figure 28). The preamps are made on boards with 0.8mm thick
aluminum cores to which the preamp operational amplifiers are cemented
using thermally conductive epoxy. The board mounts also contact this
metal core and the walls of the dewar, forming a good thermal path
from the amplifiers, which dissipate 140 mW each, to the dewar wall.
They are electrically connected with a backplane, also made of BT--resin,
which carries their signals to hermetic subminiature DB--37 connectors 

The preamp circuit diagram is shown in Figure 29.  The preamplifiers are
coupled to the output of the CCDs using a fairly standard DC
restoration circuit.  The switches in U1 are open during the video
reference and sampling time, and are closed during serial transfer time
by the logic signal IPC; the video level at sampling is thus set
approximately to I$-$TR/100, the voltage at the bottoms of the 2K
resistors R6 and R7.  The effective time constant for the circuit is the
RC time constant C1R6 divided by the duty cycle of IPC, which is 4
$\mu$s/0.16, 24 $\mu$s, or a little longer than a pixel, but there are
{\it no} AC coupling effects during the period critical to the sampling of the
video.  The mean video level is supposed to be irrelevant in a
double--correlated--sampling system, but real systems are never
perfectly balanced and it is useful to maintain tight control over the
sampling level; this circuit accomplishes this.  The amplifier itself is
a fairly fast monolithic op--amp with low--noise JFET inputs, the
Burr--Brown OPA627, run with a non--inverting gain of 10.  The gain drops
to unity for positive--going output signals (such as the reset pulse
feedthrough) greater than two diode drops by virtue of the clamp circuit
formed by D1 and D2.  With the typical output gain of 1$\mu$V/e$^-$ for
our CCDs, the preamp saturates at a signal level of 6--700,000 electrons. 
With the typical input noise level of 3.5nV/$\sqrt {\rm Hz}$ in the relevant
frequency range, the
preamp noise at our sampling rate is about 1 $\mu$V, or about one
electron.

\smallskip
\noindent
{\bf Clock drivers}\hfil

The clock drivers are four CMOS DG403 analog switches with 
RC pulse--shaping circuitry on the inputs to the CCD gates (see Figure 30). 
These switches, used basically as single--pole double--throw pairs, 
generate the CCD clocks
by attaching a CCD gate to one or another DC rail voltage supply through
suitable pulse--shaping RC circuits to limit the slew rates. The
switches used, DG403s, have switching times of order 100ns, which
is sufficiently fast for this application. (The serial overlap is
one master clock, 750ns for the slow chips and 375ns for the fast ones).
Two switches 
are used for the parallel and transfer gates, and the other two for serial
and summing well signals.
The rail voltages (VP12$\pm$, VP3$\pm$,
VT$\pm$, VS$\pm$ and VSW$\pm$) are generated by DACs on the bias
board. We have found that the CCDs are wonderfully insensitive to
the operating voltages except near the inevitable instability thresholds,
and we use voltages near the factory--recommended ones except for the
parallel levels on the UV chips, which must be run at appreciably
higher negative rails to avoid spurious charge, as remarked earlier. The
VDD levels need to be trimmed for some chips as well to get the operating
region safely away from incipient breakdown with attendant nonlinearity 
and high noise.

The astrometric preamps and clock drivers are substantially the same, but
space constraints did not allow us to use the same boards; instead, the
circuitry for twelve channels of video and the equivalent of six sets of
clock drivers is built onto one piggyback pair of boards which run the
whole length of the astrometric dewar and carry small adapter cards into
which the astrometric CCD FPCs are plugged. Some modularity is sacrificed
by this approach, but the handling and testing is really no more difficult
than with the photometric design.

\smallskip
\noindent
{\bf Signal chain board}\hfil

The signal chain electronics and bias voltage supply  are
implemented in surface--mount technology on opposite sides of
one small 8--layer 3U Eurocard. This board is mounted using a 96--pin 
DIN connector to a passive mother board mounted to the
outside of the dewar lid.

The signal--chain board handles the double--slope
integration of the CCD signal and its A/D conversion. It also
contains RS485 drivers to feed the serial A/D output to FOXI data
transmitter.  The circuitry, shown in Figure 31, consists of five
blocks. U1 (differential amplifier, gain setting, and level shifting), 
U2 (inversion) and U4 (integration) are fast low--noise Burr--Brown OPA627s and
U3 is a CMOS DG445 switch, which handles the switching of the (inverted)
reference
level during the I$-$ period and the video during the I+ period into the
input of the integrator and the reset of the integrator after the
A/D acquisition.

The input amplifier U1 operates as a differential amplifier against the
video input VID$+$ and VID$-$, which is preamp ground.  The output impedance
of the in--dewar preamplifier is 50 ohms from the op--amp and ground. 
The dewar amp output (VID+) is fed to the non--inverting input of U1. 
The gain of the amplifier is near unity, but is trimmed by
adjusting R5 and R6 for the output gain of each CCD and to set the
desired conversion gain (electrons/ADU).
This is set nominally
at 2 for the $u'$ band, 5 for $g'$ and 6 for the others.  The output level
of the CCD at about 1 $\mu$V per electron results in about a 4 volt
negative signal from the preamp at saturation, about 400,000 electrons. 

The signal chain runs with a virtual ground at the ADC reference level,
VREF$-$, $-$4.5V, to accomodate the $\pm4.5{\rm V}$ range of the ADC, with
zero output at $-$4.5V.  U1 performs the level shift using a current
injection from the VREF2 (9V) signal derived from the ADC reference. 
The video level at the output of U1 goes from $-$4.5V at zero signal to
(nominally) $-$9V at saturation.  Gains and resistor values are chosen so
that the noise contributed by all of the circuitry here is small
compared to the input noise of the preamp. 

The output of U1 goes to the input of U2, a unity--gain inverter, and
through the I$+$ switch through a 2K resistor
to the input of U4, the integrator. The reference voltage of for the
integrator is VREF$-$, which is the zero level of the ADC, and
the output of U2 is used to generate the inverted video signal for
the reference level.
Thus the integrator output goes from $-$4.5V positive; the effective gain
of the integrator is set by the 2K input resistors, the integration
capacitor C12, and the integration time, nominally 7.5 $\mu$s for the slow
chips and 3.75 $\mu$s for the fast ones. This gain, just 
$t_{integ}/(R_{in} C_{integ})$,
is set nominally to
a value near 2 with the choice of 1800 pf for C12 for the slow chips
and 910 pf for the fast ones. Thus the nominal saturation signal of $-$4.5V
at the input results in a level presented to the A/D of $+$4.5V, its
saturation level. One ADU is 130 $\mu$V; the net noise of the circuitry
on this board through the double--correlated sampling
including Johnson noise and amplifier input noise is about
23$\mu$V, much less than an ADU.

There is overload protection in the input to the integrator: 
the 2K input resistor (for the non--inverting signal) is
split into two 1K parts, R14 and R23, and the junction is clamped to a
low--impedance VREFD$-$ level by the combination of a Schottky (D1) and
an integrated 2.5V Zener (D2), and is prevented from going more
negative than $-4.5-2.5 =-7.0\rm V$, which would arise from a $-$5V input
signal. The input resistor is not connected to the input of the
integrator op amp except during the video I$+$ phase, and the integrator
end is kept from moving very far from ground
and thereby charging stray capacitance by the back--to--back 
low--leakage junction diodes (we use the base--collector diodes of signal
transistors) at the input to the switch (D6,7).
CCDs typically exhibit considerable capacitive crosstalk from the
gate (SW) which dumps the video charge onto the output node, and even if
the SW is pulsed (which we do) to dump the charge the cancellation is not
perfect. This signal 
needs to be subtracted from the video signal. This is accomplished
by an adjustable voltage (I$+$TRIM) across the 20K resistor R15.  The input
to this resistor is 10 times less noise--sensitive than the input to the
integrator, so can easily tolerate $\approx$100nV/$\sqrt{\rm Hz}$ of
noise and still contribute negligibly.  

In the double--slope integration sampling, the video level is subtracted from 
a reference level established just prior to the video dump (I$-$). The
simplest way to do this is to have the integrator integrate the negative
CCD output waveform for the same time as the positive one is integrated for the
video. The negative waveform is generated by the inverter U2, which inverts 
the output
of U1 relative to the $-$4.5V reference level. It is protected by 
a clamp consisting of a signal diode and a 5.1V Zener against overload
associated with extreme saturation. This clamp was originally implemented
with back-to-back zeners, which leaked so badly and exhibited so much
hysteresis that serious instability was introduced into the background level;
considerable sleuthing was required to find the problem, so be warned. 
(It may well be that this is a known problem, but it was certainly not known
to {\it us}.)
The output of this stage is connected to the integrator
input also through a 2K resistor (R16) and another section of the switch.

The I$-$ switch connects the output of the inverter U2 to the integrator
during the 7.5$\mu$s video sampling interval I$-$. The switch is opened
and one clock cycle elapses for the video to settle to its final value 
before the 7.5$\mu$s video sampling interval I$+$ is initiated, during 
which the output of the input amplifier U1 is connected to the integrator. 
When this period ends, the I$+$ switch is opened, and the double--slope
sampling video level is held on the output of the integrator. 
One clock cycle elapses before the conversion is begun. 

Our A/D converter is Crystal Semiconductor CS5101A,
which converts to 16 bits in 8 $\mu$s with an 8 MHz input clock 
and which has internal calibration circuitry which keeps the 
internal accuracy at $\pm$ 1/4 LSB. It requires an external 4.5V
reference and an external clock of 8.0 MHz. The timing diagram is
shown in Figure 26. The reference input must be a stable low--impedance
source. It is implemented as an OPA627 voltage follower (U102) operating
from an LT1019 reference, which also feeds the amplifiers in a quad
LT1125 to generate the other reference voltages, VREF2=9.0V, VREFD$-$=
$-$4.5V, and the critical VREF$-$=$-$4.5V, through U202a--d. 
The OPA627 has a 4.5nV/$\sqrt{\rm Hz}$ maximum input
noise, and is connected to the reference input of the amplifier via a
filter consisting of a 220$\Omega$ resistor and 11$\mu$F (C227 and 
C228) of filter 
capacitance, the DC loop being closed around the filter. The noise
bandwidth is less than 1 kHz, and the noise injected into the reference
input is less than 1$\mu$V.

There is some
concern about whether 16 bits is enough, but the problem is not severe.  
In the $g'$, $r'$, $i'$, and $z'$
bands the sky levels are such that digitizing at roughly 6 electrons/(AD unit)
results in a step which is small compared to the standard deviation in
the background, and accommodates with 16 bits the nominal 300,000 electron
full well of the device.  In the $u'$ chips the situation is not quite
so good.  The sky in each chip is only of order 60 electrons, and the noise
with 5 electrons read noise per chip is about 9 electrons.  If we demand that we
digitize at at least two levels per unit sigma, this will result in a
digital full well of somewhat less than the physical full well of
the CCD.  This is not a serious problem, however, since the fluxes and
efficiencies are such that no object which does not saturate in one of
the other bands is likely to overflow the $u'$ A/D. In practice we
also back off a bit for the $g'$ chips to an inverse gain of 5e$^-$/ADU.

In addition to the problem with the Zener clamp on the inverting
amplifier, we had another serious problem with this board which
manifested itself in rather troublesome levels of 1/f noise in the
video.  This was not present in the test data taken with much older
electronics, so the CCDs themselves seemed to be absolved.  The
culprits, again after much investigation, turned out to be {\it
resistors}.  We had been careful to use metal-film resistors in all the
gain-determining places on the boards, but had used ordinary precision
thick-film surface-mount resistors elsewhere.  These turn out to have
completely unacceptable current noise in 16-bit systems; replacing the
feedback components and voltage divider resistors in the reference
circuit with metal-film ones (which fortunately can be had in the small
0603 SMT format for which the boards were designed) finally solved the
problem.  If we had the boards to redo from scratch, we would use
metal-film resistors everywhere. 

The signal--chain board dissipates about 1.5 watts; there is one per chip
in the photometric dewars, one per two devices (all single--amplifier) in the
astrometric dewars.

\smallskip
\noindent
{\bf Serial Shift register and FOXI board}\hfil

The Crystal A/Ds have a serial output, which is fed (again over RS485)
to a pair of small cards mounted piggy--back on the dewars whose function 
is to format the data and transmit it over fiber to the data system
in the operations building. In the photometric dewars, there is typically
a mixture of single-- and two--amplifier chips, and the fast serial data
from the single--amplifier devices must be formated to look like two
streams so that it is compatible in timing and data rate with the output 
of the normal chips. This is accomplished by the serial shift--register card,
which processes the single--amplifier data and passes it on along with the
unprocessed two--amplifier data to the second card, which carries the 
fiber transmitter.

The FOXI card carries a Xilinx programmable logic array which can receive
up to 6 double--amplifier serial data streams with their associated clocks,
produces sequential parallel byte--wide data, and sends it to a FOXI fiber
transmitter, which reserializes the data at about 100 MHz and drives 
the data fiber.

The astrometric dewars have two FOXI boards with slightly different
firmware and no serial shift--register boards; though there are only 12
single--amp data channels, the data rates are such that two transmitter
boards are required.

\smallskip
\noindent
{\bf Bias board}\hfil

The bias board generates the DC voltages necessary for 
operation of CCDs and their 
associated linear support electronics (except the ADC and associated 
signal chain references, which are independent; there is no good reason
for this, and it may well have been a mistake; it would probably have
been better to derive all voltages from the ADC reference).
In addition, there are two more functions of the bias board, {\it viz},
it acts as an analog multiplexer to allow measurement of those
voltages by a remote ADC incorporated into the camera microprocessor,
and it provides temperature control for CCDs. The circuitry is shown
in Figure 32. 

The CCD voltages and the voltage required to trim the dual--slope 
circuitry are generated by a set of three octal 8--bit DACs, Maxim 528s,
for each chip. 
These voltages provide rail potentials for CCD gates which have small
(a few tens of pf) capacitances (T, SW, R) or are nearly constant
current loads (VDD). Simple filters and slow operational amplifiers
suffice to keep the voltage nearly constant. For instance, for the
transfer gate (T) the gate capacitance is $\approx 100$ pf and switched
every 40 ms through a range of about 10 V. At every transition the charge 
on the filter cap changes by 1000 pC, so the voltage across 11$\mu$f
of filters drop by 100 $\mu$V.  With a 200 $\Omega$ series resistor,
the voltage is reestablished with a time constant of about 2 ms,
sufficiently short compared to the line time. 

Another example of a low-capacitance gate but one with very high
sensitivity to voltage errors is the
summing well. It has 10pf capacitance and switched every 22.5 $\mu$s.
It has a two--stage passive filter with a time constant of about 3 ms,
and the average current 4.2 $\mu$A.
The total filter resistance is 1200 $\Omega$, so the drop is 5 $\mu$V.
The summing well is an especially sensitive case for CCDs, since feedthrough
from the summing well transition emerges directly in the video output:
while all other clock feedthrough is differentiated by the double slope
sampling, the SW clock dumps the charge directly on the output diffusion.
This feedthrough amounts to a signal corresponding to 15000 electrons of 
image charge. If therefore one wishes a system controlled to 1 electron,
it is necessary to keep the rail voltage of the SW switch stable to of
the order of 1/15000, or 0.8 mV for the switching range of 12 V. The
voltage drop of our system is 5 mV, but the summing well runs all 
the time, so it is stable to the order of resistor stability, clock
stability, and reference stability; there are no timing glitches in
the reading sequence of the CCD which interrupt it.  In practice, we
{\it pulse} the summing well to dump the charge, which to some approximation
cancels the feedthrough, but the stability requirements still apply.  

The serial drives have about 1000 pf capacitance per phase with another
1000 pf buffer on the dewar board, and each phase is switched every
22.5 $\mu$s, with a given transition followed by another in 1.5 $\mu$s.
The switching time occupies 5.25 $\mu$s of the 22.5 $\mu$s the serial
read time. The switching amplitude is typically 12 volts, and the switching
current about 3 mA. With the individual gate time constant 120 ns,
the instantaneous switching currents are large, about 200 mA. At each
switching transition 25 nC of charge is removed from the 22 $\mu$f
filters, and so the voltage changes by about 1 mV. If we must recover
from this by the next transition from the same rail 1.5 $\mu$ s later,
it would require a current from the regulator of about 16 mA with 
response time shorter than 2$\mu$s. This could be done with
high speed op amps, but there is really no need to respond this fast: 
the three transitions from one rail changes the voltage by only 3 mV in
6 $\mu$s, tiny compared to the total swing and small enough to keep
the op amp out of slew, and the regulator then has 14 $\mu$s to
recover. 

The parallel drives pose a similar problem. The parallel gate capacitance
is very high, 80 nf per phase. There is 70 nf of buffer on the dewar
board, and so the total switched capacitance is about 150 nf.
Each parallel clock is 8 pixels long, or 180  $\mu$s. The
waveform is controlled by the clock capacitance, buffer and an external
series RC circuit (400 $\Omega$, 40--100 $\Omega$ from effective gate
resistance and 300--360 $\Omega$ of serial buffer resistance). 
With a voltage swing
of 12 V, the peak current is about 30 mA. The switch transitions to a
given rail are about 360  $\mu$s, and two transitions
follow by 37 ms of silence, during which the parallel voltages must be
very stable.  The gates cover so much of the device that potential changes
show up in the output as gain changes in the output amplifier. Here we
have elected to allow the regulator op amps to go nonlinear. A transition
subtracts about 1.8  $\mu$C of charge from the 22  $\mu$f filters with
a step of about 82 mV. The regulators, U2 and U11, are directly connected
to the filters. This voltage step drives the op amps into current slew,
and they deliver their short circuit current, 25 mA for LT1124, until
the deficit is made up about 60  $\mu$s after.  There is a glitch of
about 10  $\mu$s until the amplifier settles, but the voltage is well
stabilized before the next transition occurs, and very well settled 
after the last transition in the 360  $\mu$s before the serial readout
starts.  

The regulators, U3, for VDD1 and VDD2 used in the output amplifier incorporate
some current protection. The LM334 current sources act like forward--biased
diodes so long as the current through them is less than their set currents,
3 mA, but go to a very high impedance state as the current approaches the
set value. The nominal output FET drain current is 1 mA, and the op amp 
divider and measurement divider require another 0.7 mA. For these currents
the resistance of LM334 is of the order of 100 $\Omega$; the high--frequency
impedance is reduced by the 0.68 $\mu$f shunt capacitor. Another 0.68 $\mu$f
shunt to the inverting input limits the noise output to input voltage
noise of the op amp, which is 18 nV/$\sqrt {\rm Hz}$. 

The DACs have serial input, and all the converters in the system are
daisy--chained to be set up with one very long serial word.  The output
voltages are sampled and placed on an output bus for measurement with a
set of serial--input 8--position analog switches, U19,U20, and U21. 
Twenty--three of the switches are used for voltages to be monitored, and
the last to connect or disconnect the selected voltage from the output
line from the board.  This last enables a crude bus architecture in
which one line going to the ADC on the executive microprocessor can
monitor all the voltages in the camera.  The steering logic is contained
on the bus receiver board. 

The default voltage information is stored in EPROMs in the controller
micro, but can be modified in software.

The voltages generated by the DACs are monitored by a 10--bit ADC which
is built into the Hitachi H8 Forth microcontroller which runs the
camera.  The voltages are switched one by one onto a dewar--wide voltage
bus by the DG486 serial--input octal switch arrays, and
thence buffered and switched onto a camera--wide bus by circuitry on each
dewar's bus receiver board.  The time constants are such that the camera
can measure accurately one voltage per five line times during the
operation, with the setup happening during a 900 $\mu$s parallel
transfer time. Thus the system can run gathering one voltage value per
130 ms.  There are 24 voltages for each of 30 photometric and 12 pairs
of astrometric CCDs, so the process requires about 130 seconds for one
cycle.  For all voltages which have passive filters the measurements are
taken ahead of the filters to avoid any disturbances to the operating
voltages. 

The bias board also contains the temperature regulator for the
chip it serves.
The CCD temperature sensors are platinum resistors which have 
a resistance of 1.00k$\Omega$ at 0 C. They have a characteristic
temperature dependence of resistance which is almost proportional
to the absolute temperature. The slight nonlinearity in this relationship
can be compensated approximately by feeding the platinum resistor 
with a stable voltage source through a suitably chosen (relatively large)
stable resistor and using the voltage across the platinum resistor
as the temperature signal. This resistor is the 17.3K unit R50.
The first stage of the regulator amplifier U16cd is a gain 25 non--inverting
amplifier, whose output is $+$50mV/C and is biased so that its output
is zero at 0 C. The zero can be trimmed with the DAC--generated level
TZERO. The output is compared with another level, TSET, by the very
high--gain second stage, which drives the emitter--follower pass
transistors located on the power distribution board which supply
current to the makeup heaters on the CCDs. TSET determines the
threshold temperature at which the resistors turn on, and the gain
is such that the transition from full OFF to full ON is over a range
of about a degree. Not having an abrupt transition appears to be
necessary to avoid hunting in the temperature servo, and the
accuracy and stability with this gain is satisfactory.

There is one of these cards per chip on the photometric camera. Some
of these cards, associated with the single--amplifier chips, use only
one signal channel, but all the cards are identical except for the input
resistors which are used to trim the gains to match the individual output
amplifier gains for each CCD, and the integrator capacitor, which is
a different value for the fast single--amplifier circuitry. 
Thus each photometric dewar
contains 5 of these cards, plus one card which is basically a bus
receiver for the clock signals described below; the
astrometrics employ 6 plus a bus receiver. All astrometric
chips run with only one amplifier (they are frontside devices and will
all run fast) to save cards and space, 
and so we run two chips with each signal--chain--bias card---the performance
penalty is negligible, and the astrometric chips are sufficiently
homogeneous in electrical (and other) properties that the loss of ability
to generate voltages tailored to a specific CCD is of no concern; in
any case, we retain the ability to set the output drain voltage and
reference voltage for each chip.

Each bias board consumes altogether 1.3 W, of which about 0.5 W is associated
with DACs. Since the bias board is physically associated with a signal
chain board, each serves one device in the photometric dewars and two
in the astrometric.

\smallskip
\noindent
{\bf Bus receiver}\hfil

The bus receiver is a modified small Eurocard which converts the
RS485 timing signals on the 100--pin ribbon cable bus to CMOS logic
levels which are fed to the rest of the camera circuitry. It also
contains a shift register and circuitry which decodes the last byte of the
monitor voltage--selection word. The voltage--selection word is four
bytes long, the first three of which have one bit set among them to
select the voltage one wishes to connect to the executive micro's ADC
and the last of which contains the dewar and CCD address. The dewar
ID is set with jumpers on the board, and if the dewar address agrees
with the dewar ID, the output of the buffer amplifier U12 is connected
to the bus. The CCD address is decoded and translated into an enable
signal for each CCD's signal/bias board, which routes the serial selection
bits to the relevant dewar via the multiplexer U9. The switches on the
bias boards come up with all switches cleared; the procedure is then to
send a setting word of all zeros to the last dewar/CCD addressed to
turn off the switch for the last voltage measured, send a byte to connect
the new one, and send a word to connect the desired voltage for that
dewar/CCD. When it has been measured, the sequence is repeated.
Figure 33 shows the circuitry.

\smallskip
\noindent
{\bf The Camera Controller}\hfil

The camera controller is built around two Hitachi H8/532 FORTH microprocessors
from
Triangle Data Services.  One, the {\it executive micro}, manages the
housekeeping and communications, and the other, the {\it phase micro},
generates the CCD clock waveforms.  There is also support circuitry
required to implement the various housekeeping functions and bus
interfaces.  Each micro is mounted piggyback on a small Eurocard and
there are in addition five other Eurocards which are part of the
controller system: a {\it bus driver} which drives the RS485 CCD timing
bus and implements the check byte comparator for the long serial
DAC--setting word; a {\it star driver} which generates the master 8MHz
clock, derives the 2.3333 MHz pixel clock from it and has the 24 RS485
drivers to send the master clock to all the dewars and all the FOXI
transmitters; an {\it LN2 board}, which implements all the autofill
logic for the dewars; a {\it pressure board}, an electropneumatic
nightmare which handles the bellows pressure/vacuum for the Tbar
latches, monitors those pressures, and handles the interface to the
telescope control system which commands the latches; and a {\it
temperature/shutter board} which controls the electropneumatic shutters
and buffers all the various other pressures, temperatures, and miscellaneous
states which the
executive micro monitors.  These include air temperatures in the camera
and in the power supplies, the temperature of the steel main support
ring for the camera, pressures in the intermediate LN2 supply dewars, and
the state of a flow switch in the water/glycol cooling system for the
electronics. 
The analog signals are multiplexed by this card to send to the executive
micro ADC, and the digital ones buffered for sending to a port on that
micro.

The phase micro generates the CCD clocks from programs stored in EPROMs and
controlled by an Hitachi H8/523 Forth microcontroller.
One controller runs the entire camera; all CCDs
are clocked synchronously with identical clocking waveforms (though
there are two sets of serial waveforms for the `fast' and `slow' chips.)
The system diagram is shown in Figure 34.

The executive micro has an (optical fiber) RS232 connection which is
one of the two communications connections to the outside world. It
controls the phase micro with another RS232 connection. The Hitachi
microprocessor has an integral 4--channel 10--bit A/D which is used to measure
the CCD voltages placed on the voltage bus on one channel and to
measure the pressures, temperatures, etc, in the housekeeping data
with another. The CCD voltages (the CCD temperatures are part of this
set) are measured on a round--robin basis 
as described above. The DAC setting requires a word a little more
than 10000 bits long, which has a leading check--byte which is
received back at the bus driver after making the entire rounds.  The
clock is slow, 10kHz, and so the setup time is about 1 second.  All
signals are sent from the bus driver by balanced RS485 transmitters and
received at each dewar by complementary receivers.  All signals except
the 8 MHz master clock are carried on a 100--conductor ribbon cable bus;
each 8 MHz line, one to each dewar and one to each FOXI card, 
has a dedicated shielded twisted pair cable from the star driver.

The executive has considerable flexibility in controlling the
chips; we implement on--chip binning both
horizontally and vertically (independently) and allow some control of
the sampling interval. Several clocking schemes are stored in ROM,
including one in which the video switch I+ is never turned on to
allow setting the I--TRM video levels, one in which there is no
data sampling to allow quick access to any row in the chip for partial
readout, etc.

\smallskip
\noindent
{\bf The Power System}\hfil

Each dewar has its own power supply, each of which is implemented using
small encapsulated linear (dictated by noise considerations) supplies on
a single $20\times25$ cm 3 mm thick printed--circuit card.  Each supply
carries a fairly sophisticated monitoring and shutdown system mounted on
a separate smaller PC card.  A set of LEDs and a digital meter allows
one to read any selected voltage, display the status of all voltages
(in--range or out, selected or not for the meter).  A sufficiently
out--of--range voltage shuts down the whole supply, but the culprit is
still displayed on the status display. An error insufficient to cause alarm
but still out-of-spec (the thresholds are typically set at 3 percent and 5
percent of the desired value) illuminates the out-of-range LED for that
voltage but does not cause a shutdown; in this way drifts can often 
be caught before shutdown occurs. The power supply boards are mounted
in their own enclosures mounted on a structure called the {\it saddle}
(see Section 8) which stays with but is not mechanically tied to the
camera.

The DC power lines go from the supply chassis through the camera enclosure
and thence through DB-25 connectors on the individual dewars to a
{\it power distribution board}. This is a small board which carries filters for
each supply, regulators to make the voltages used by the preamps 
($\pm 10{\rm V}, 5{\rm V}$), the pass transistors for the CCD heaters,
and the amplifier for the platinum resistor temperature sensor on the 
LN2 container. Power for the FOXI boards is carried directly to them and
is not, for noise reasons, routed through the dewar electronics.

\medskip
\noindent
{\bf 8. The Overall Structure of the Camera and its Utility Support}\hfil

The fused quartz distortion corrector provides the metering structure for the
detectors and allows us to keep the relative geometry within the focal
plane extremely stable, but it is, of course, necessary to hold the
corrector stably with respect to the telescope as well. This is complicated
by the fact that the corrector has a very small expansion coefficient,
more than an order of magnitude smaller than the steel structure of the
telescope. We decided to take up the expansion difference in the design
of the corrector cell, and to build the main camera structure of steel.

The corrector is mounted in a thin machined steel cylindrical cell which
is sliced vertically to form thirty ``leaves'' which are bonded to the
circumference of the corrector.  These leaves are quite flexible in the
radial direction owing to their small thickness (1/8 inch) but very stiff in
shear tangentially to the corrector.  This ring is, in turn, bolted to a
very stiff steel ring girder which forms the main structural element of
the camera (see Figure 35).

The ring girder, welded from two short 1/4 inch steel tubes and two 3/8 inch
steel rings and then machined as one piece, has an approximately 3 x 7 inch
cross section and, with the cell, weighs about 200 pounds.  This ring
carries all the mass associated with the camera except the power
supplies and the intermediate LN2 dewars; this ``camera proper'' mass is
about 700 pounds.  The ring is mounted to the instrument rotator bearing on
the telescope on a fairly conventional classical trefoil kinematic mount,
implemented in this case as three sets of double roller bearings
allowing free radial movement (see Figure 36).  
Each leaf of the trefoil utilizes eight
rollers, four each defining a surface parallel to the radius in one
direction and 45\deg from the optical axis in the other. 
The kinematic mount is maintained in contact by three pneumatically
actuated sprung latches, one associated with each leaf, each exerting a
force of 700 pounds along the axis of the camera.  This arrangement,
combined with the leaved cell for the corrector, results in deflections
under gravity which are very small and are furthermore very well
behaved.  The corrector center moves by about 5$\mu$m with respect to
the fixed (telescope) members of the kinematic mounts from the zenith to
the horizon, and that deflection is to high accuracy in the direction of
the local gravity vector, independent of the rotational orientation
of the kinematic mount, and is accompanied by completely negligible twist.
The focus change is of order 2 microns. Both the focus deflection and
the lateral deflection are completely negligible compared to expected 
deflection effects in the telescope structure and optics.

A pneumatically--operated shutter assembly is attached to the top of the
leaved corrector cell (Figure 37). This utilizes 16 hinged doors, two
associated with each dewar.  The opening/closing times are of the order
of 1 second, and it is intended primarily as a light seal, though
crudely timed exposures can be made with it.

The camera is covered on top by a welded aluminum ring structure
attached to the main ring girder which serves to carry a dust cover when
the camera is stowed and offers protection to the shutter assembly and
corrector. Flexible rubber gaskets seal this cover to the inside of the
instrument rotator when the camera is on the telescope. This seal 
keeps stray light out and also serves as a crude seal for a continuous
nitrogen purge between the camera and the Gascoigne corrector of the
telescope, about 70cm above the distortion corrector. This purge is
primarily to keep the optics clean.

A similar shroud is attached to the bottom.  This is a considerably more
complex structure. It carries all the connectors for power and
signals, the cooling system for the electronics, and the LN2 plumbing
which carries nitrogen from the intermediate dewars to the camera dewars
and the vent gas from the camera dewars to the solenoid vent valves;
these valves are also mounted on this ring.  A conical cover is attached to the
bottom of the ring and serves as a light--tight cover for the camera. 
The dewars are surrounded on the corrector by a machined aluminum light
shield as well, the intent being that the space behind the corrector
which the dewars occupy should be reasonably sealed both to gas and
light, and certainly that no light from the telescope enclosure should
reach the focal plane from the dewar side of the corrector.  This space
is also kept under nitrogen purge, the primary objective here being to
keep the dewars and their associated cryogenic lines dry. A photograph
of the dewars with their associated liquid nitrogen plumbing and cooling
vent tubes in the rear shroud before all the wiring and insulation were
added is shown in Figure 38.

The dewar power supply boards are distributed in two
almost identical large chasses, each carrying four dewar supplies and an
``auxiliary'' supply for the microprocessors and associated control
electronics, the fans, and the valves.  These are mounted along with the
two 10--liter intermediate LN2 dewars on an aluminum saddle (Figure 39)
which always accompanies the camera but is mounted to the telescope
independently of it. This is in order that the kinematic mount 
carries only what
it must, and that as symmetric and stiff as possible.  The saddle
assembly with power supplies and dewars
weighs about 300 pounds, so the camera assembly all together
weighs just over 1000 pounds. 

The camera and saddle are carried when off the telescope on a special
cart/handling fixture (the {\it ops cart}) which rolls on fixed rails
between the telescope and the ``doghouse'', a sealed enclosure which
houses the camera when it is not in use. A photograph of the camera
alone on the telescope with the saddle on the ops cart in the doghouse
is shown in Figure 40.  This unusual configuration is the one used to
transfer the camera to the service cart (see below) for transport
for servicing.  AC Power, gas utilities,
glycol/water coolant, signal fibers, and liquid nitrogen are carried to
the camera on an umbilical which always remains attached to the camera
on or off the telescope.  The camera will be kept cold continuously
except for times of maintenance.  For serious maintenance and repairs,
there is another cart (the {\it service cart}) which allows the camera
to be taken to the clean room in the operations building, and inverted
to reach and remove the dewars (Figure 41).  
This cart carries the camera alone; the
saddle remains on the ops cart, though for full operation in the clean
room (which is possible) the intermediate dewars and power supplies have
to be removed and transported with the camera. 

The camera consumes about 550 watts when quiescent, as much as 100
more at peak. The AC power is supplied by a 1 KVA motor--generator
through an isolating UPS. The heat dissipated is disposed of
by heat exchangers, two in the camera body and three in each
power supply chassis. Each power supply chassis dissipates about 160 watts
and the camera electronics about 220, though the LN2 system in the
camera disposes of more than half the latter. The heat exchangers
are supplied with water/glycol at about 4C below ambient by a chiller
operating below the observing floor, and the instrument structure is
thereby kept near ambient temperature.

The electronics in the photometric dewars dissipate about 25 watts each
and those for the astrometrics about 32.  One of the heat exchangers in
the camera is coupled to a set of hoses through which air is blown
directly into the electronics enclosures on the dewars (see Figure 37); 
the other 
simply circulates the air in the rear camera shroud.

The communication with the outside world is reasonably simple for so
complex an instrument.  There are sixteen data fibers, 10 for CCD data
from the camera FOXIs, two for the RS232 control signals, and 4 for the
control of the Tbar latches by the telescope control system; AC power;
water/glycol for cooling (4.5 liters/min) and its return; low-pressure
gaseous nitrogen
for purge and pressurization of the intermediate dewars; 
vacuum and pressurized nitrogen for the dewar bellows; 
compressed air for the shutters; and a vacuum--jacketed liquid nitrogen line
for filling the intermediate dewars.  All of these are carried in the
umbilical, a bundle 2.5 inches in diameter which is covered and protected
by a sturdy flexible ethylene vinyl acetate hose with steel wire reinforcement.

\noindent
\medskip
{\bf 9. Conclusion}\hfil

We have described in this paper the construction of a large multi--CCD
camera to be used in the Sloan Digital Sky Survey.  The camera is
currently at the Apache Point site.  It has undergone extensive testing,
including mechanical fit--checks, mating to the data system, cold
system--level tests, and final shakedown.  As of this writing it has
been on the sky three times, once for checkout near full moon and then two
weeks later, in late May 1998, for an official `first-light' run in the
dark of the moon, and again in the next dark run in June. 
Though there have been the usual minor commissioning problems, no major
problems have surfaced, and the data from the June dark run,
taken when the telescope was much better collimated than in the May run, 
is quite good.
The project has been both
long and arduous, though very rewarding.  We hope this rather detailed
description will be of assistance to those building large CCD systems in
the future. 

\noindent
\medskip
{\bf 10. Acknowledgements}\hfil

The camera team would like to thank the participating institutions in
the SDSS, including Princeton University, the University of Chicago, the
Institute for Advanced Study, the
Japanese Participation Group, Fermilab, the University of Washington,
the Naval Observatory, and the Johns Hopkins University for support during
often difficult times, and of course the Alfred P.  Sloan Foundation and
the National Science Foundation for the grants which have allowed the
survey to proceed.  A Grant-in-Aid for Specially Promoted Research,
No. 05101002, from the Ministry of Education in Japan was awarded
specifically for some aspects of the construction of this camera, and
the Japanese team was supported by the Japan Society for the Promotion
of Science.  We also thank especially Keith Gollust for a generous gift
of both money and faith which allowed the project to begin in a serious
way seven years ago, and the PI on the camera wishes to thank Barry and
Bobbi Freedman and John and Karen Nichols for help at critical times which made
working on the camera possible.
\vfil
\eject
\parindent=0pc
{\bf References}

\ref Abe, F., 1997, in {\it Variable Stars and the Astrophysical Returns of 
    Microlensing Surveys}, R. Ferlet, J. P. Maillard and B. Raban, eds.,
    Fronti\`eres, Paris, 75.

\ref Abe, F., \etal, 1998, in preparation.

\ref Arnaud, M., Aubourg, G., Bareyre, P., Brehin, S., Caridroit, R., 
    deKat, J., Dispau, G., Djidi, K., Gros, M., \& Lachieze-Rey, M., 
    1994, Experimental Astronomy 4, 265.

\ref Bauer F., \& J. deKat (for the EROS2 collaboration), 1998,
    in {\it Optical Detectors for Astronomy}, J. Beletic and P. Amico, eds.,
    Kluwer, Amsterdam, 191.

\ref Boroson, T., Reed, R., Wong, W., Lesser, M., 1994, Proc. SPIE 2198, 877.

\ref Boulade, O., 1998, in {\it Optical Detectors for Astronomy}, J. Beletic 
    and P. Amico, eds.,  Kluwer, Amsterdam, 203.

\ref Bowen, I. S. \& Vaughan, A. H. 1973, Applied Optics, 12, 1430.

\ref Fukugita, M., Ichikawa, T., Gunn, J. E., Doi, M. Shimasaku, K. \&
     Schneider, D. P. 1996, AJ, 111, 1748

\ref Gunn, J. E. \& Westphal, J. A. 1981, Proc. SPIE, 290, 16

\ref Gunn, J. E., Carr, M., Danielson, G. E., Lorenz, E. O., Lucinio, R.,
    Nenow, V. E., Smith, J. D. \& Westphal, J. A. 1987, Optical Eng., 26, 779.

\ref Ives, D., Tulloch, S., Churchill, J., 1996, Proc. SPIE 2654, 266.

\ref James, E., Cowley, D., Faber, S., Hillyard, D., \& Osborne, J.,
    1998, Proc. SPIE 3355, 70.

\ref Kashikawa, N., Yagi, M., Sekiguchi, M., Okamura, S., Doi, M., 
    Shimasaku, K., \& Yasuda, N., 1995, IAU Symp 167.

\ref Luppino, G., Metzger, M., Kaiser, N., Clowe, D., Gioia, I., \&
    Miyazaki, S., 1996, in ASP Conf. Ser. 88, {\it Clusters, Lensing, and
    the Future of the Universe}, V. Trimble, ed., ASP San Francisco, 229.

\ref Miyazaki, S., 1998, Proc. SPIE 3355, in press.

\ref Schneider, D. P., Gunn, J. E. \& Hoessel, 1983, ApJ, 264, 337.

\ref Schneider, D. P., Schmidt, M. \& Gunn, J. E. 1989, AJ 98, 1951.

\ref Sekiguchi, M., Iwashita, H., Doi, M., Kashikawa, N., \& Okamura, S., 
    1992, PASP 104, 744.

\ref Thuan, T. X. \& Gunn, J. E. 1976, PASP, 88, 543.

\ref Turnrose, B, 1974, PASP, 86, 545.

\ref Wittman, D., Tyson, J., Bernstein, G., Lee, R., Dell'Antonio, I.,
    Fischer, P., Smith, D., \& Blouke, M., 1998, Proc. SPIE 3355, 626.

\vfil\eject
\vfil
$$\vbox{
\rm
\noindent
\tabskip=0.6em plus 1em minus 0.5em
\centerline{Table 1. Summary of TDI images with scan scale 3.6343 mm/arcmin}
\halign to \hsize{
\hfil  #  & \hfil  #  & \hfil  # & 
\hfil  #  & \hfil  #  & \hfil  # &
\hfil  #  & \hfil  #  & \hfil  # &
\hfil  #  & \hfil  #  & \hfil  # \cr
\noalign{\vskip 3mm \hrule\vskip 0.1mm}
\noalign{\vskip 1mm \hrule\vskip 3mm}
field & $x$ & $y$ & CCD size & fil & ffc3 & ccd4 & vscale &
dfoc & dc4 & em & eM \cr
& mm  & mm  &  mm$\times$mm & & mm$^{-1}$ & mm$^{-1}$ & mm/$'$ & $\mu$m &
mm$^{-1}$ & $\mu$m & $\mu$m \cr
\noalign{\vskip 2mm \hrule\vskip 2mm}
 1&  45.5&    0.0& 49.2$\times$ 49.2& $u'$& 1.37e-3& 4.3e-4& -3.6317&  5&  0.3e-4& 17& 35\cr
 2& 136.5&    0.0& 49.2$\times$ 49.2& $u'$& 1.37e-3& 4.3e-4& -3.6332&  8& -0.3e-4& 19& 25\cr
 3& 227.5&    0.0& 49.2$\times$ 49.2& $u'$& 1.37e-3& 4.3e-4& -3.6335& 28& -1.3e-4& 30& 38\cr
 4&  45.5&   65.0& 49.2$\times$ 49.2& $i'$& 1.40e-3& 4.3e-4& -3.6350&  4&  0.2e-4& 18& 21\cr
 5& 136.5&   65.0& 49.2$\times$ 49.2& $i'$& 1.40e-3& 4.3e-4& -3.6353& 13& -0.4e-4& 20& 24\cr
 6& 227.5&   65.0& 49.2$\times$ 49.2& $i'$& 1.40e-3& 4.3e-4& -3.6339& 30& -1.3e-4& 25& 35\cr
 7&  45.5&  -65.0& 49.2$\times$ 49.2& $z'$& 1.40e-3& 4.3e-4& -3.6348&  4&  0.1e-4& 18& 21\cr
 8& 136.5&  -65.0& 49.2$\times$ 49.2& $z'$& 1.40e-3& 4.3e-4& -3.6350& 13& -0.5e-4& 20& 25\cr
 9& 227.5&  -65.0& 49.2$\times$ 49.2& $z'$& 1.50e-3& 4.3e-4& -3.6347& 26& -1.1e-4& 28& 39\cr
10&  45.5&  130.0& 49.2$\times$ 49.2& $r'$& 1.30e-3& 4.3e-4& -3.6342& 12& -0.5e-4& 18& 19\cr
11& 136.5&  130.0& 49.2$\times$ 49.2& $r'$& 1.30e-3& 4.3e-4& -3.6338& 25& -1.0e-4& 20& 23\cr
12& 227.5&  130.0& 49.2$\times$ 49.2& $r'$& 1.50e-3& 4.3e-4& -3.6347& 37& -1.4e-4& 26& 32\cr
13&  45.5& -130.0& 49.2$\times$ 49.2& $g'$& 1.30e-3& 4.3e-4& -3.6350& 10& -0.4e-4& 19& 22\cr
14& 136.5& -130.0& 49.2$\times$ 49.2& $g'$& 1.30e-3& 4.3e-4& -3.6347& 23& -0.9e-4& 21& 23\cr
15& 227.5& -130.0& 49.2$\times$ 49.2& $g'$& 1.40e-3& 4.3e-4& -3.6350& 39& -1.5e-4& 27& 31\cr
16&  45.5&  204.5& 49.2$\times$ 9.6&  $r'$& 1.50e-3& 4.3e-4& -3.6354&  2& -0.1e-4& 19& 21\cr
17& 136.5&  204.5& 49.2$\times$ 9.6&  $r'$& 1.50e-3& 4.3e-4& -3.6350& 10& -0.8e-4& 22& 26\cr
18& 227.5&  204.5& 49.2$\times$ 9.6&  $r'$& 1.80e-3& 4.3e-4& -3.6362& 20& -1.4e-4& 30& 40\cr
19&   0.0&  220.0& 49.2$\times$ 9.6&  $r'$& 1.50e-3& 4.3e-4& -3.6353&  1& -0.0e-4& 20& 21\cr
20&  91.0&  220.0& 49.2$\times$ 9.6&  $r'$& 1.50e-3& 4.3e-4& -3.6352&  6& -0.4e-4& 21& 24\cr
21& 182.0&  220.0& 49.2$\times$ 9.6&  $r'$& 1.60e-3& 4.3e-4& -3.6355& 17& -1.2e-4& 26& 33\cr
22&   0.0&  235.5& 49.2$\times$ 9.6&  $r'$& 1.50e-3& 4.3e-4& -3.6353&  2& -0.2e-4& 22& 23\cr
\noalign{\vskip 2mm \hrule\vskip 2mm}
}}$$
\vfil
\eject
\null
\vfil
$$\vbox{
\rm
\vskip2cm
\noindent
\tabskip=3mm plus 1mm minus 1mm
\centerline{Table 2. CCD noise and QE}
\halign to \hsize{
\hfil  # \hfil & \hfil  # \hfil & \hfil  # \hfil & 
\hfil  # \hfil & \hfil  # \hfil \cr
\noalign{\vskip 3mm \hrule \vskip 0.1mm }
\noalign{\vskip 1mm \hrule \vskip 3mm}
filter& max noise($e^-$) & $<{\rm QE}>$ & $\lambda$(\AA) &$e^-$ in sky \cr
\noalign{\vskip 2mm \hrule \vskip 2mm }
 $u'$&       7&               0.36&   3500&        45\cr
 $g'$&       7&               0.73&   4700&       400\cr
 $r'$&       9&               0.82&   6400&       700\cr
 $i'$&      20&               0.69&   7700&      1200\cr
 $z'$&      16&               0.18&   9200&      1100\cr
\noalign{\vskip 2mm \hrule\vskip 2mm}
}}$$
\vfil
$$\vbox{
\rm
\noindent
\centerline{Table 3. SDSS Filter Characteristics and Photometric Sensitivity, 1.4 Airmasses}
\tabskip=3em plus 1em minus 2em
\halign to \hsize{
 # \hfil &  # \hfil  &  # \hfil &  # \hfil &  # \hfil \cr
\noalign{\vskip 3mm \hrule\vskip 0.1mm}
\noalign{\vskip 1mm \hrule\vskip 3mm}
filter & $\lambda_{\rm eff}$  & FWHM & $ q_t$ & $Q$ \cr
\noalign{\vskip 3mm \hrule\vskip 3mm}
$u'$ &  3549  &  560&  0.111 & 0.0116 \cr
$g'$ &  4774  & 1377&  0.436 & 0.113  \cr
$r'$ &  6231  & 1371&  0.549 & 0.114  \cr 
$i'$ &  7615  & 1510&  0.490 & 0.0824 \cr
$z'$ &  9132  &  940&  0.128 & 0.0182 \cr
%
\noalign{\vskip 3mm \hrule \vskip 3mm}
}}$$
\vfil
\eject
\null
\vfil
$$\vbox{
\rm
\noindent
\centerline{Table 4. Photometric Saturation and Sky Background}
\tabskip=3mm plus 1mm minus 1mm
\halign to \hsize{
# \hfil & # \hfil & # \hfil & # \hfil & # \hfil & # \hfil \cr
\noalign{\vskip 3mm \hrule\vskip 0.1mm}
\noalign{\vskip 1mm \hrule\vskip 3mm}
Filters                &   $u'$&   $g'$&   $r'$&   $i'$&   $z'$\cr
\noalign{\vskip 2mm \hrule\vskip 2mm}
Star saturates at AB   &   12.0&   14.1&   14.1&   13.8&   12.3\cr
Eff Sky, mag/sec$^2$   &   22.1&   21.8&   21.2&   20.3&   18.6\cr 
Sky + bkg count/pxl    &     40&    390&    670&   1110&   1090\cr
\noalign{\vskip 3mm \hrule \vskip 3mm}
}}$$
\vfil
\eject
\null
\vfil
$$\vbox{
\rm
\noindent
\baselineskip=\normalbaselineskip
\centerline{Table 5. Photometric Sensitivity for V$_{sky}=21.7$, $\sec z = 1.4$}
\tabskip=1em plus 1em minus 0.5em
\halign to \hsize{
 # \hfil & \hfil # & \hfil # & \hfil # & \hfil # & \hfil # &
\hfil # & \hfil # & \hfil # & \hfil # & \hfil # \cr
\noalign{\vskip 3mm \hrule\vskip 0.1mm}
\noalign{\vskip 1mm \hrule\vskip 3mm}
filter && $u'$  && $g'$ && $r'$ && $i'$ && $z'$ \cr
\noalign{\vskip 3mm \hrule\vskip 3mm}
AB & count & S/N & count & S/N & count & S/N & count & S/N & count & S/N \cr
\noalign{\vskip 3mm \hrule\vskip 3mm}
17.0& 28033& 160.4& 194863& 428.2& 194863& 420.2& 140849& 338.2& 31281& 124.2\cr
17.5& 17688& 124.5& 122950& 334.4& 122950& 325.0&  88869& 255.0& 19737&  86.7\cr
18.0& 11160&  95.5&  77577& 258.8&  77577& 248.1&  56073& 188.3& 12453&  59.0\cr
18.5&  7042&  72.1&  48947& 197.9&  48947& 186.1&  35380& 135.7&  7857&  39.3\cr
19.0&  4443&  53.3&  30884& 148.7&  30884& 136.6&  22323&  95.2&  4958&  25.7\cr
19.5&  2803&  38.5&  19486& 109.4&  19486&  97.8&  14085&  65.2&  3128&  16.7\cr
20.0&  1769&  27.1&  12295&  78.5&  12295&  68.2&   8887&  43.6&  1974&  10.7\cr
20.5&  1116&  18.6&   7758&  54.8&   7758&  46.4&   5607&  28.7&  1245&   6.8\cr
21.0&   704&  12.4&   4895&  37.4&   4895&  30.9&   3538&  18.6&   786&   4.3\cr
21.5&   444&   8.2&   3088&  24.9&   3088&  20.2&   2232&  12.0&   496&   2.7\cr
22.0&   280&   5.3&   1949&  16.4&   1949&  13.1&   1408&   7.6&   313&   1.7\cr
22.5&   177&   3.4&   1230&  10.6&   1230&   8.4&    889&   4.9&   197&   1.1\cr
23.0&   112&   2.2&    776&   6.8&    776&   5.4&    561&   3.1&   125&   0.7\cr
23.5&    70&   1.4&    489&   4.3&    489&   3.4&    354&   1.9&    79&   0.4\cr
24.0&    44&   0.9&    309&   2.8&    309&   2.2&    223&   1.2&    50&   0.3\cr
24.5&    28&   0.6&    195&   1.7&    195&   1.4&    141&   0.8&    31&   0.2\cr
25.0&    18&   0.4&    123&   1.1&    123&   0.9&     89&   0.5&    20&   0.1\cr
\noalign{\vskip 3mm \hrule\vskip 3mm}
}}$$
\vfil
\eject
\null
\vfil
$$\vbox{
\rm
\noindent
\centerline{Table 6. The Circuit Board Complement}
\tabskip=2em plus 2em minus 1em
\halign to \hsize{
 # \hfil &  \hfil #  &  \hfil  # \cr
\noalign{\vskip 3mm \hrule\vskip 0.1mm}
\noalign{\vskip 1mm \hrule\vskip 3mm}
parts     &  photometric &   astrometric \cr
\noalign{\vskip 3mm \hrule \vskip 3mm}
CCD   &  30 & 24 \cr
photo preamp  &  30 &  \cr 
photo clk driver    & 30  &  \cr
astro preamp & & 2\cr
astro clk driver  & & 2\cr
signal/bias & 30 & 12 \cr
bus receiver & 6 & 2 \cr
power regulator board & 6 & 2 \cr
serial shift register & 6 & 0 \cr
FOXI board & 6 & 4 \cr
\noalign{\vskip 3mm \hrule \vskip 3mm}
}}$$
There are, in addition, eight unique boards associated with the camera 
controller, eight each of two different boards in the power supplies,
and a board which controls the intermediate dewar automatic fill.
\vfil\eject
%
\noindent
{\bf Figure Captions}

\item{\rm Fig. 1.} The optical layout of the focal plane of the SDSS camera.
Field 22 (top and bottom) are focus CCDs; fields 16-21 are astrometric
chips, and 1-15 are the photometric array. The TDI scan direction is
upward, so a star traverses the array from top to bottom.

\item{\rm Fig. 2.} Greyscale images of the PSF in the distinct optical
fields identified in Figure 1.  The bottom panel shows the output of the
optical design with no seeing; the top images are those convolved with
0.8 arcsecond FWHM gaussian seeing.  The five images for each field are
for five positions horizontally across the chip, evenly spaced from one
extreme edge to the other. 

\item{\rm Fig. 3.} Quantum efficiencies of the three varieties of CCDs
used in the camera, showing for the thinned Vis-AR coated chips a
representative range of variability encountered among different devices
of the same type.  Note that the Vis-AR curves are for room temperature,
the others cold.  At operating temperature the infrared response of the
Vis-AR devices is comparable to the others.  On the scale of this figure
it is not possible to see easily the extended infrared response of the
front-side devices. 

\item{\rm Fig. 4.} An image of a Ronchi ruling at a peak exposure level
of 30 electrons.  The period of the ruling is about 6 pixels, and the
image is taken from a single--amplifier frame from the corner farthest
from the amplifier.  Defects associated with a small number of vertical
traps can be seen in the image, the most prominent one and its
associated head slightly above the center near the left edge of the
image. 

\item{\rm Fig. 5.} The system quantum efficiency for each filter/detector
system in the photometric array. The expected throughput of the optics
is included; the lower of each pair of curves includes the expected 
atmospheric extinction associated with 1.2 airmasses at our site.

\item{\rm Fig. 6.} The layout of the six photometric and eight
astrometric dewars in the camera. The photometric dewars interlock
with each other on the mounting rails (see Figure 8), and the astrometric
dewars interlock with the ends of the photometric ones. Note the three
filters associated with each focus device.

\item{\rm Fig. 7.} Photograph of a disassembled photometric dewar at an early
stage of construction. The tubes which carry pressurized nitrogen to the
kinematic mount bellows can be seen, as well as the backplane circuit
board which interconnects the in-vacuum preamp/clock--driver boards with
the hermetic connectors which carry the signals into the vacuum. The
CCD mountings on the Tbar can be seen, as well as the copper cold posts which
carry heat from the CCDs to the cold strap assembly.

\item{\rm Fig. 8.} Photograph of the front surface of the corrector showing
the filters cemented onto the back face. The striped antireflection
coating can be seen in the foreground, as well as a hint from the reflection
of the ceiling fluorescent light tubes of the extremely aspheric surface
of the element. The focus chip filters had not been glued on at the time
this photograph was taken.

\item{\rm Fig. 9.} Photograph of the rear surface of the corrector before
the filters were cemented on, showing the dewar mounting rails and the
quartz pillars for the Tbar kinematic mounts. Again the
reflections of the ceiling fluorescents (the ring-shaped feature in the
upper right) show the peculiarity of the front figure.

\item{\rm Fig. 10.}  A Pro/Engineer CAD drawing of the CCD on its mount,
here displayed as a cutaway. The ball--and--cone socket nature of the
mount can be seen as well as the Kovar stiffener to which the CCD is
cemented.

\item{\rm Fig. 11.}  A Pro/Engineer CAD representation of an exploded
view of the Tbar assembly. Refer to the photographs in Figures 7 and 12.

\item{\rm Fig. 12.}  Photograph of the photometric Tbar assembly, showing
the Tbar itself and a set of ball-and-socket mounts. Note the four
tilt adjusting screws in the front socket. The Kovar stiffener/ball mounts
have the copper cold posts attached, and the Tbar has its kinematic
ends mounted with the upper set of rods for the kinematic mounts in place.

\item{\rm Fig. 13.} A Pro/Engineer exploded view of a Tbar kinematic mount
and latch. The bellows have Delrin extension posts which engage the conical
holes in the latch plate and pivot the plate down to disengage the latch
hook as they extend.

\item{\rm Fig. 14.} Photograph of bellows and springs which pull the
Tbar away from the kinematic mounts when the bellows are relaxed. 
Missing is the small spring which keeps the latch engaged then.  The
Tbar is raised by the springs and engages conical stops mounted on the
thin-walled tubular stainless-steel tripod structure when the bellows
are relaxed.  It is held reasonably securely there, but the camera must
be vertical for the latches to operate satisfactorily. 

\item{\rm Fig. 15.} The stage of the XYZ measuring machine used to set
up the tilt and rotation and verify the focus of the photometric CCDs on
the Tbars.  A Tbar can be seen mounted on the carriage of the linear
rail just in front of the microscope illuminator. 

\item{\rm Fig. 16.} Cutaway drawing of a photometric dewar, showing
the layout of the CCDs and kinematic mounts, and that of the circuit
boards, FOXI transmitter, and LN2 container.

\item{\rm Fig. 17.} Photograph of circuit boards under test on a
photometric dewar lid, showing power cable, CCD clock bus ribbon, 8 MHz
line, and top boards.  A set of power supply boards can be seen in the
background. 

\item{\rm Fig. 18.}  Response of astrometric detector/filter system 
(light curve) scaled up by a factor of 10 and compared with the photometric
$r'$ response (heavy curve). The extension to the red of the astrometric
passband is clearly seen. Not included in the scaling is the effect
of the factor of five shorter exposure time owing to the reduced extent
of the astrometric CCDs in the scan direction.

\item{\rm Fig. 19.} Photograph of astrometric CCDs on a partially assembled
astrometric Tbar. The focus sensor and bridge chips are mounted. The
FPCs and light shields mounted on the CCDs can be seen.

\item{\rm Fig. 20.} Pro/Engineer drawing of an astrometric Tbar, showing
all CCDs, the astrometric kinematic mounts, and the cooling assembly.

\item{\rm Fig. 21.} Images at the focus sensor thru focus, as produced by
the optics alone on the right and convolved with 0.8 arcsecond FWHM gaussian
seeing on the left. The five images
horizontally are at five locations across the chip. The center images
vertically are in focus, and progressively out of focus in 150 $\mu$m
steps in both directions.

\item{\rm Fig. 22.} Photograph of the LN2 container used in both the
photometric and astrometric dewar. It is a vacuum-brazed assembly
of copper with welded stainless--steel plumbing. It holds 400 ml of
liquid nitrogen. The Swagelok fittings at the near end mate to the
supply and vent lines; the post at the far end mates to the cold-strap
assembly shown in Figure 23.
\item{\rm Fig. 23.} Photograph of the cold-strap assembly for a photometric
dewar. The clamp at the top mates to the cylindrical post on the LN2 can,
and the blocks at the bottom attach to the copper cold posts on the
CCD mounts (see Figures 7, 11, and 12.) The leaves which attach these
blocks to the heavy copper spreader bar are made of multiple layers of
2-mil thick silver and are very flexible.

\item{\rm Fig. 24.} Conceptual diagram of the wiring of a photometric
dewar, showing the main signal pathways and functions of the various
circuit boards. See Figures 16 and 17 for the actual physical layout.

\item{\rm Fig. 25.} The timing diagram for the parallel transfer of
charge. The time unit is eight complete (slow) pixel times, and the horizontal
clocks except for the serial gates and the start-convert signal
(see figure 26) continue to run during the vertical sequence to minimize
line-start transients. Note that this parallel sequence is the same for
all devices, slow and fast.

\item{\rm Fig. 26.} The timing diagram for the serial transfer of charge
and the signal processing control for one-- and two-- amplifier CCDs.

\item{\rm Fig. 27.} Photograph of the socket board assembly on a CCD
mount and the preamp/clock--driver module, showing the FPC used to
carry the CCD signals to the electronics.

\item{\rm Fig. 28.} Sketch showing how a preamplifier is mounted to the dewar
wall, how it plugs in to connectors on the backplane board, and how
the backplane board is connected to the motherboards on the dewar lids
through the hermetic connectors. Note that it is the in-vacuum backplane
board which is soldered to the connectors epoxied into to the lid and is
hence nonremovable; the motherboards on top can be removed with care.

\item{\rm Fig. 29.} Schematic diagram of one channel of the preamplifier.

\item{\rm Fig. 30.} Schematic diagram of the clock driver circuitry.

\item{\rm Fig. 31.} Schematic diagram of the signal chain, showing the
reference voltage generators and one channel of signal processing and
analog--to--digital conversion.

\item{\rm Fig. 32.} Schematic diagram of the bias board.

\item{\rm Fig. 33.} Schematic diagram of representative circuits on the
bus receiver board, detailing the voltage monitor selection and 
monitoring circuitry and giving examples of the RS485 handling. The
long serial chain for setting the bias DACs is handled using the ribbon
cable clock bus by alternately cutting the conductors on the extreme
edges of the cable and setting the jumpers detailed here accordingly.

\item{\rm Fig. 34.} Conceptual diagram of the camera controller showing the
major signal paths.

\item{\rm Fig. 35.} Pro/Engineer cutaway of the main camera ring and
corrector in its cell. The corrector is bonded into the thin-wall slotted
steel cell, and bending of the ``leaves'' in the cell takes up the
expansion difference between the steel and quartz.

\item{\rm Fig. 36.} Pro/Engineer exploded view of one of the three kinematic
mounts which mate the camera to the telescope. The male and female blocks
are hardened EDM-cut steel. The spacer shown below the female block, which
is the side mounted to the camera, is made of Ultem 2300 resin, a strong
plastic which has an expansion coefficient well matched to steel, and serves
as part of a system which keeps the camera electrically isolated from the
telescope structure.

\item{\rm Fig. 37.} Photograph of the top shroud of the camera through
the opening in which can be seen the shutter assembly, with the
``French door'' photometric shutters. Less obvious at the top are the
astrometric shutter doors, also shown open in this photograph.

\item{\rm Fig. 38.} Photograph of the camera during assembly, showing
the microprocessor box at the extreme right, the photometric dewar
tops, the dewar vacuum manifold (here attached to a turbomolecular
pump through the flex hose exiting upper left), the liquid nitrogen
supply and return tubing (before the application of foam insulation),
and the air cooling lines to the photometric dewars at the bottom.

\item{\rm Fig. 39.} Pro/Engineer drawing of the camera with the front and
rear shrouds removed and the saddle assembly as it is mounted relative
to the camera. The two 10-liter intermediate supply dewars are shown
mounted on the saddle, but the power supplies, which are mounted on the
saddle faces adjacent to the dewars, are not shown here. Also shown
is the umbilicus, the kinematic mounts, and one of the pneumatic
latches (left center) which hold the camera to the telescope.

\item{\rm Fig. 40.} Photograph of the camera mounted alone on the
telescope, with the saddle and operations cart in the doghouse
in the background.

\item{\rm Fig. 41.} The camera inverted on the transport/service cart
in the Apache Point clean room. It is shown here fully assembled 
except for the back cover, which is easily removed for access and 
for pumping the dewars, which is in progress here. The XYZ measuring
engine can be seen in the upper right background.

%
\vfill
\eject
\bye